\documentclass[%
pre
,twocolumn%
,superscriptaddress
,amsmath,amssymb,aps,nobibnotes,
nofootinbib
,preprintnumbers
]{revtex4-1}

\usepackage{hyperref}
\usepackage[dvipdfmx]{graphicx}
\usepackage{bm}
\usepackage{braket}
\usepackage{color}
\usepackage{tabularx}
\usepackage{mathtools}

\newcommand{\bp}{\overline{\psi}}
\newcommand{\br}{{\bf r}}
\newcommand{\rhoa}{\rho_{\rm v}}
\newcommand{\rhoi}{\rho}

\begin{document}
\title{Functional-renormalization-group approach to classical liquids with short-range repulsion: a scheme without repulsive reference system}

\author{Takeru Yokota}
\email{takeru.yokota@riken.jp}
\affiliation{Interdisciplinary Theoretical and Mathematical Sciences Program (iTHEMS), RIKEN, Wako, Saitama 351-0198, Japan}
\affiliation{Institute for Solid State Physics, The University of Tokyo, Kashiwa, Chiba 277-8581, Japan}
\author{Jun Haruyama}
\email{haruyama@issp.u-tokyo.ac.jp}
\affiliation{Institute for Solid State Physics, The University of Tokyo, Kashiwa, Chiba 277-8581, Japan}
\author{Osamu Sugino}
\email{sugino@issp.u-tokyo.ac.jp}
\affiliation{Institute for Solid State Physics, The University of Tokyo, Kashiwa, Chiba 277-8581, Japan}
\date{\today}


\begin{abstract}
The renormalization-group approaches for classical liquids in previous works
require a repulsive reference such as a hard-core 
one
when applied to systems with short-range repulsion.
The need for the reference is circumvented here by using a functional renormalization group approach for integrating the hierarchical flow of correlation functions along a path of variable interatomic coupling. We introduce the cavity distribution functions to avoid the appearance of divergent terms and choose a path to reduce the error caused by the decomposition of higher order correlation functions. We demonstrate using an exactly solvable one-dimensional models that the resulting scheme yields accurate thermodynamic properties and interatomic distribution at various densities when compared to integral-equation methods such as the hypernetted chain and 
the Percus-Yevick equation, even in the case where our hierarchical equations are truncated with the Kirkwood superposition approximation, which is valid for low-density cases.
\end{abstract}

\maketitle

\section{Introduction\label{sec:intro}}
In the context of the statistical-mechanical theory 
for classical liquids,
there is a long history for the studies of integral equations
governing density correlation functions, or distribution functions \cite{han13}.
A famous and successful one is that relying on the Ornstein--Zernike (OZ) equation
with approximated closures, 
such as the hypernetted chain (HNC) and 
the Percus--Yevick (PY) equation.
How to systematically improve the accuracy in this framework
is, however, still an open problem.
The Bogoliubov--Born--Green--Kirkwood--Yvon (BBGKY) hierarchy of equations 
is another well-known and rigorous framework.
It is, however, still a challenging problem to describe dense systems 
accurately with this hierarchy.
For instance, precise knowledge about the lower- and higher-order distribution functions
beyond the Kirkwood superposition approximation (KSA) 
is required to describe dense systems
with the BBGKY hierarchy of equations
as, truncated with KSA,
it might magnify the error induced by KSA \cite{bar76}
and actually
shows poor results at high densities \cite{kir50,lev66}
in comparison with HNC and PY.

The renormalization group (RG) is another fundamental notion 
to capture properties of many-body systems,
where differential equations associated with the scale transformation
play the central role.
The concept of RG has also been applied to the analysis of classical liquids,
see, e.g., Refs.~\cite{par85,par93,par95,par08,par09,par12,sal92,whi93,whi95,iso19,cai06,cai11,lue15} 
and the references therein, which include
the hierarchical reference theory (HRT) \cite{par85,par93,par95,par08,par09,par12}
known as a combination of RG and thermodynamic perturbation theory,
the application of Wilson's phase-space cell method \cite{sal92,whi93,whi95},
and RG with respect to the scale transformation of density \cite{iso19}.

An established framework for RG is
the functional renormalization group (FRG) \cite{weg73,wil74,pol84,wet93} (for reviews, see, e.g., Refs.~\cite{ber02,paw07,met12,dup21}),
in which the one-parameter evolution of the system
is described by an exact differential equation, which is
called a flow equation, for some functional.
The formalism based on the effective action \cite{wet93}
being the counterpart of the bare action incorporating the thermal and quantum fluctuations
is a sophisticated framework
as the flow equation is described by a closed functional differential
equation for the effective action,
which provides systematic ways to analyze many-body systems
incorporating non-perturbative effects.
There are several works for the application to classical liquids \cite{cai06,cai11,lue15},
where FRG becomes a framework to treat the free-energy density functional $F[\rho]$
of the particle-number density $\rho(\br)$,
which corresponds to the effective action multiplied by the temperature.
In Refs.~\cite{cai06,cai11}, 
formal aspects of FRG for classical liquids
and some analytic results are presented. 
Some numerical demonstration in the case of 
the gradient expansion employed as the approximation
are shown in Ref.~\cite{lue15}.

FRG for the calculation of density functionals,
i.e., FRG formulated for density functional theory (DFT),
has also been developed in the case of quantum many-body systems,
as initiated in Refs.~\cite{pol02,sch04}.
In this direction, some numerical applications and formal extensions,
which include numerical analyses of low-dimensional toy models
\cite{kem13,ren15,kem17a,lia18,yok18,yok18b}
and two- and three-dimensional electron systems \cite{yok19,yok21}
and extension to the case of superfluid systems \cite{yok20},
have been recently achieved.
The approximations employed in these works are on the basis of 
the vertex expansion,
where the functional Taylor expansion is employed and
the functional flow equation is converted to a hierarchy of flow equations
for density correlation functions.
These studies suggest the possibility of FRG to 
actually contribute the improvement of accuracy of DFT.

RG approaches for classical liquids developed until now including FRG
provides various ways to incorporate the effect of long-range weak force.
The contribution from short-range repulsive force is, 
however, usually treated in an empirical and less systematic manner:
Most of the works rely on the approach in which 
a reference system is employed to incorporate the contribution of the short-range repulsion.
This approach requires the knowledge of the reference system
and causes dependence of the results on the empirical choice of the reference system,
albeit being successful when choosing a well-studied and well-behaved reference as suggested in the study of HNC \cite{sum16}.

In this paper, we develop an FRG formulation for classical liquids, 
where the free energy density functional is evolved along a path of variable interatomic coupling. 
Herein the hierarchical flow of correlation functions obtained from the vertex expansion
is stabilized for a system having short-range repulsive forces by introducing cavity distribution functions 
instead of reference system representing the contribution of short-range repulsion. 
After discussing appropriate choice for the path of variable coupling, we demonstrate the performance of our approach using one-dimensional exactly solvable models. 
In terms of the thermodynamic properties and interatomic distribution at high densities as well as low densities, our scheme formulated at the level of KSA for the higher order correlation functions is already superior to integral-equation methods such as HNC and PY.

This paper is organized as follows:
In Sec.~\ref{sec:flo},
we derive the flow equation for 
the free-energy density functional
and the hierarchy of equations
for the cavity distribution functions.
The discussion about a suitable choice
of evolution for short-range repulsion is also in there.
Section \ref{sec:demo} shows
the numerical demonstration of our method
in one-dimensional models.
Section \ref{sec:con} is devoted to
the conclusion.
In Appendix \ref{sec:derivation_flow},
the details of the derivation of our hierarchy of equations
for distribution functions are described.

\section{FRG and hierarchy of equations\label{sec:flo}}
We first summarize the formalism of the FRG 
for classical many-body systems.
In this paper, we restrict our discussion to the case of two-body interaction 
and consider its evolution.
The evolving two-body interaction is denoted by
$v_{\lambda}(\br-\br')$, where $\lambda$ is the evolution parameter running from $0$ to $1$
and $v_{\lambda}(\br-\br')$ continuously changes with respect to $\lambda$
satisfying the boundary condition
$v_{\lambda}(\br-\br')$ as $v_{\lambda=1}(\br-\br')=v(\br-\br')$ being 
the two-body interaction of interest.
As a way to treat strong repulsive part it is possible to set the repulsive part as the initial condition $v_{\lambda=0}(\br-\br')$, 
but we do not introduce a reference for the repulsive part
and set $v_{\lambda=0}(\br-\br')=0$.
There is a freedom for the choice of the evolution toward $v_{\lambda=1}(\br-\br')$.
The appropriate choice may depend on the model for $v(\br-\br')$ and approximation scheme.
Let us leave the discussion about it until Sec.~\ref{sec:evolution_hc}
and focus on the derivation of the flow equation.

Let $\mu$ and $U(\br)$ be the chemical potential and external field, respectively.
Since these quantities always appear in the form of
$\beta(\mu-U(\br))$ with the inverse temperature $\beta$,
we introduce the notation 
$\bp(\br) \coloneqq \beta(\mu-U(\br))$.
As a functional of $\bp(\br)$, 
the grand partition functional is given by
\begin{align}
	\label{eq:def_partition_fcn}
	\Xi_{\lambda}[\bp]
	\coloneqq
	&
	\sum_{N=0}^{\infty}
	\frac{1}{\Lambda^{3N} N!}
	\int_{\br_1,\ldots,\br_N}
	e^{
	-\sum_{i<j}^{N}\beta v_\lambda(\br_i-\br_j)
	+\sum_{i=1}^{N}\bp(\br_i)
	},
\end{align}
where we have introduced the short-hand notation
$\int_{\br_1,\ldots,\br_N}
=\int d\br_1 \cdots \int d\br_N$
and the de Broglie thermal wavelength $\Lambda$.
From this grand partition functional, the thermodynamic potential is given by
\begin{align}
	\label{eq:def_omega}
	\Omega_\lambda[\bp]
	\coloneqq
	-\frac{1}{\beta} \ln \Xi_{\lambda}[\bp],
\end{align}
which plays the role of the generating functional for the density correlation functions.
The Helmholtz free energy is defined by the Legendre transformation of $\Omega_\lambda[\bp]$:
\begin{align}
	\label{eq:def_helm}
	F_\lambda[\rhoa]
	&\coloneqq
	\max_{\bp}
	\left[
	\Omega_\lambda[\bp]
	+
	\frac{1}{\beta}
	\int_{\br}
	\rhoa(\br)\bp(\br)	
	\right]
	\notag
	\\
	&=
	\Omega_\lambda[\bp_{\max,\lambda}[\rhoa]]
	+
	\frac{1}{\beta}
	\int_{\br}
	\rhoa(\br)\bp_{\max,\lambda}[\rhoa](\br),
\end{align}
where the variable $\rhoa(\br)$
stands for the particle-number density and
$\bp_{\max,\lambda}[\rhoa]$ is given through
\begin{align}
	\label{eq:def_psimax}
	-\left.
	\frac{\delta \beta\Omega_\lambda[\bp]}{\delta \bp(\br)}
	\right|_{\bp=\bp_{\max,\lambda}[\rhoa]}
	=
	\rhoa(\br),
\end{align}
and satisfies
\begin{align}
	\label{eq:F_psimax}
	\frac{\delta \beta F_{\lambda}[\rhoa]}{\delta \rhoa(\br)}
	=
	\bp_{\max,\lambda}[\rhoa](\br).
\end{align}
These relations suggest that $\bp_{\max,\lambda}[\rhoa](\br)$ is the external field
giving the density $\rhoa(\br)$.
Conversely speaking, for a given external field 
$\bp_{\rm given}(\br)$, $\rhoa(\br)$ is determined from
$\bp_{\max,\lambda}[\rhoa](\br)=\bp_{\rm given}(\br)$.

\subsection{Flow equation}
\begin{figure}[!htb]
  \begin{center}
    \includegraphics[width=0.6\columnwidth]{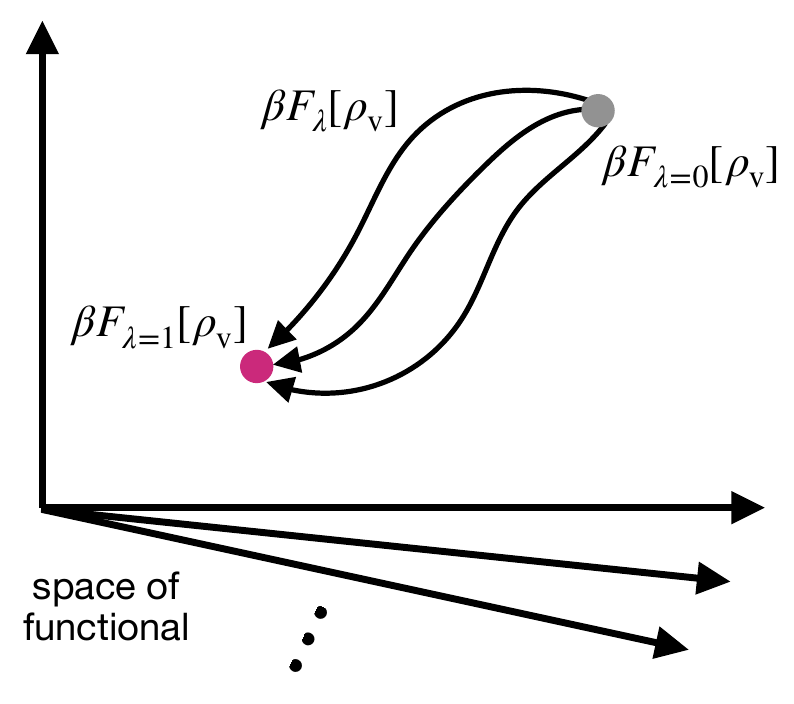}
    \caption{Schematic picture of the evolution of $\beta F_\lambda[\rhoa]$.
    The different paths correspond to different choices of $\beta v_{\lambda}(\br-\br')$.}
    \label{fig:flow_eq}
  \end{center}
\end{figure}

The evolution of $F_\lambda[\rhoa]$ with respect to $\lambda$
can be described by a flow equation in a closed form for $\beta F_\lambda[\rhoa]$
as follows:
\begin{widetext}
\begin{align}
	\label{eq:Fflow_eq}
	\partial_\lambda
	\beta F_\lambda[\rhoa]
	=&
	\frac{1}{2}
	\int_{\br,\br'}
	\partial_\lambda
	\beta v_{\lambda}(\br-\br')
	\left(
	\rhoa(\br)\rhoa(\br')
	+
	\left(
	\frac{\delta^2 \beta F_{\lambda}}{\delta \rhoa\delta \rhoa}
	\right)^{-1}[\rhoa](\br,\br')
	-
	\rhoa(\br)\delta(\br-\br')
	\right),
\end{align}
\end{widetext}
which we will derive shortly.
Following this flow equation,
$\beta F_\lambda[\rhoa]$ evolves in the functional space
as illustrated in Fig.~\ref{fig:flow_eq},
where the path of the evolution depends on the choice of $\beta v_{\lambda}(\br-\br')$.
A similar form of the flow equation
is known for quantum cases \cite{sch04,kem13,kem17a,yok18,yok19}.

The derivation of Eq.~\eqref{eq:Fflow_eq} is as follows:
The derivative of Eq.~\eqref{eq:def_helm}
with respect to $\lambda$ together with
Eq.~\eqref{eq:def_psimax} leads to
\begin{align}
	\label{eq:dFlam}
	\partial_\lambda
	\beta F_\lambda[\rhoa]
	=
	(\partial_\lambda \beta \Omega_{\lambda})[\bp_{\max,\lambda}[\rhoa]].
\end{align}
Note that the partial derivative on the right-hand side 
does not act on the argument of
$\Omega_{\lambda}$, i.e., $\bp_{\max,\lambda}[\rhoa]$.
By use of the definitions
Eqs.~\eqref{eq:def_partition_fcn} and \eqref{eq:def_omega},
Eq.~\eqref{eq:dFlam} is rewritten as
\begin{widetext}
\begin{align}
	\label{eq:flow_omega}
	\partial_\lambda
	\beta\Omega_\lambda[\bp_{\max,\lambda}[\rhoa]]
	=&
	\frac{1}{\Xi_\lambda[\bp_{\max,\lambda}[\rhoa]]}
	\sum_{N=0}^{\infty}
	\frac{1}{\Lambda^{3N}N!}
	\int_{\br_1,\cdots,\br_N}
	\sum_{i<j}^{N} 
	\partial_\lambda
	\beta v_{\lambda}(\br_{i}-\br_{j})
	e^{
	-\sum_{i'<j'}^{N}
	\beta v_{\lambda}(\br_{i'}-\br_{j'})
	+
	\sum_{i'=1}^{N}\bp_{\max,\lambda}[\rhoa](\br_{i'})
	}
	\notag
	\\
	=&
	\int_{\br,\br'}
	\partial_\lambda
	\beta v_{\lambda}(\br-\br')
	\left
	\langle
	\sum_{i<j}^{\hat{N}}
	\delta(\br-\hat\br_{i})
	\delta(\br'-\hat\br_{j})
	\right
	\rangle_{\lambda,\bp_{\max,\lambda}[\rhoa]}
	\notag
	\\
	=&
	\frac{1}{2}
	\int_{\br,\br'}
	\partial_\lambda
	\beta v_{\lambda}(\br-\br')
	\left
	\langle
	\hat{\rho}(\br)\hat{\rho}(\br')
	-
	\hat{\rho}(\br)\delta(\br-\hat\br')
	\right
	\rangle_{\lambda,\bp_{\max,\lambda}[\rhoa]},
\end{align}
where
\begin{align}
	\label{eq:def_ave}
	\left\langle
	\cdots
	\right\rangle_{\lambda,\bp}
	=&
	\frac{1}{\Xi_\lambda[\bp]}
	\sum_{N=0}^{\infty}
	\frac{1}{\Lambda^{3N} N!}
	\int_{\br_1,\cdots,\br_N}
	\cdots
	e^{
	-\sum_{k<l}^{N}\beta v_\lambda(\br_{k}-\br_{l})
	+\sum_{k=1}^{N}\bp(\br_{k})
	},
\end{align}
and
$\hat{\rho}(\br)=\sum_{i=1}^{\hat{N}}\delta(\hat\br_i-\br)$
with $\hat{N}$ and $\hat\br$ being variables to be averaged.
From Eqs.~\eqref{eq:def_partition_fcn}, \eqref{eq:def_omega},
and \eqref{eq:def_psimax}, one finds
\begin{align}
	\label{eq:r1}
	\left
	\langle
	\hat{\rho}(\br)
	\right
	\rangle_{\lambda,\bp_{\max,\lambda}[\rhoa]}
	=&
	\rhoa(\br),
	\\
	\label{eq:r2}
	\left
	\langle
	\hat{\rho}(\br)
	\hat{\rho}(\br')
	\right
	\rangle_{\lambda,\bp_{\max,\lambda}[\rhoa]}
	=&
	\rhoa(\br)\rhoa(\br')
	-
	\left.
	\frac{\delta^2 \beta\Omega_\lambda[\bp]}{\delta \bp(\br)\delta \bp(\br')}
	\right|_{\bp=\bp_{\max,\lambda}[\rhoa]}.
\end{align}
As shown from the derivative of Eq.~\eqref{eq:def_psimax}
with respect to $\bp_{\max,\lambda}(\bf r)$
and 
the derivative of Eq.~\eqref{eq:F_psimax}
with respect to $\rhoa(\br)$,
the following inverse relation holds:
\begin{align}
	\label{eq:finv}
	\left(
	\frac{\delta^2 \beta F_{\lambda}}{\delta \rhoa\delta \rhoa}
	\right)^{-1}[\rhoa](\br,\br')
	=
	-
	\left.
	\frac{\delta^2 \beta\Omega_\lambda[\bp]}{\delta \bp(\br)\delta \bp(\br')}
	\right|_{\bp=\bp_{\max,\lambda}[\rhoa]},
\end{align}
where the left-hand side is defined through
\begin{align}
	\int_{\br''}
	\left(
	\frac{\delta^2 \beta F_{\lambda}}{\delta \rhoa\delta \rhoa}
	\right)^{-1}[\rhoa](\br,\br'')
	\frac{\delta^2 \beta F_{\lambda}[\rhoa]}{\delta \rhoa(\br'')\delta \rhoa(\br')}	
	=
	\delta(\br-\br').
\end{align}
From Eq.~\eqref{eq:dFlam} together with Eqs.~\eqref{eq:flow_omega} and
\eqref{eq:r1}-\eqref{eq:finv}, we finally obtain
Eq.~\eqref{eq:Fflow_eq}.
\end{widetext}


\subsection{Vertex expansion}
In principle, $\beta F_\lambda[\rhoa]$,
and thus all the thermodynamic quantities and correlation functions,
are obtained by solving Eq.~\eqref{eq:Fflow_eq} only.
It is, however, nontrivial whether
Eq.~\eqref{eq:Fflow_eq} can be used practically.
In particular, the functional equation is hard to treat numerically
in a direct manner as the space of the argument $\rhoa(\br)$ 
is computationally huge, and introduction of some approximation is practically needed.
Here, we employ the vertex expansion,
in which the functional Taylor expansion around
\begin{align}
	\rhoa(\br)=\rhoi(\br)
\end{align}
with $\rhoi(\br)$ being some density of interest, 
and truncate the expansion at some order to realize numerical calculation.
Since the derivative coefficients of $\beta F_\lambda[\rhoa]$ are related to
correlation functions, the hierarchy of flow equations can be written
in terms of the correlation functions. Let us introduce the $m$-particle density:
\begin{align}
	\label{eq:rn_def}
	&\rho^{(m)}_\lambda(\br_1,\ldots,\br_m)
	=
	\left\langle
	\prod_{i=1}^{m}
	\left(
	\hat{\rho}(\br_i)
	-
	\sum_{j=1}^{i-1}\delta(\br_j-\br_i)
	\right)		
	\right\rangle_{\lambda,\bp_{\max,\lambda}[\rhoi]}.
\end{align}
Through the derivations as summarized
in Appendix~\ref{sec:derivation_flow}, 
the following
hierarchy of the flow equations
are obtained:
\begin{widetext}
\begin{align}
	\label{eq:flow_Fn}
	&\partial_\lambda
	\beta F_\lambda
	=
	\frac{1}{2}
	\int_{\br,\br'}
	\partial_\lambda
	\beta v_{\lambda}(\br-\br')
	\rho^{(2)}_\lambda(\br,\br'),
	\\
	\label{eq:flow_rhom}
	&\partial_\lambda
	\rho_\lambda^{(m)}(\br_1,\ldots,\br_m)
	+
	\sum_{i<j}^{m}
	\partial_\lambda
	\left[\beta v_\lambda(\br_i-\br_j)\right]
	\rho_\lambda^{(m)}(\br_1,\ldots,\br_m)
	\notag
	\\
	&
	\quad
	=
	\int_{\br}
	\left[
	\rho^{(m+1)}_\lambda(\br_1,\ldots,\br_m,\br)
	-
	\rho(\br)
	\rho^{(m)}_\lambda(\br_1,\ldots,\br_m)
	\right]
	\partial_\lambda \bp_{\max,\lambda}(\br)
	\notag
	\\
	&
	\qquad
	+
	\rho^{(m)}_\lambda(\br_1,\ldots,\br_m)
	\sum_{i=1}^m
	\partial_\lambda \bp_{\max,\lambda}(\br_i)
	-
	\int_{\br}
	\sum_{i=1}^m
	\partial_\lambda
	\left[\beta v_\lambda(\br-\br_i)\right]
	\rho^{(m+1)}_\lambda(\br, \br_1, \ldots, \br_m)
	\notag	
	\\
	&
	\qquad
	-
	\frac{1}{2}
	\int_{\br,\br'}
	\partial_\lambda\left[\beta v_\lambda(\br-\br')\right]
	\left[
	\rho^{(m+2)}_\lambda(\br,\br',\br_1,\ldots,\br_m)
	-
	\rho^{(2)}_\lambda(\br,\br')
	\rho^{(m)}_\lambda(\br_1,\ldots,\br_m)
	\right],
\end{align}
\end{widetext}
with $m\geq 1$,
$\beta F_\lambda \coloneqq \beta F_\lambda[\rho]$, and 
$\bp_{\max,\lambda}(\br) \coloneqq \bp_{\max,\lambda}[\rho](\br)$.
Note that in the case of $m=1$,
Eq.~\eqref{eq:flow_rhom} is regarded as the flow equation
for $\bp_{\max,\lambda}(\br)$ rather than that for $\rho^{(1)}_\lambda(\br)$
since we fix $\rhoi({\br})$ during the flow and $\rho^{(1)}_\lambda(\br)$ always satisfies
$\partial_\lambda\rho^{(1)}_\lambda(\br)=\partial_\lambda \rho(\br)=0$.

\subsection{Cavity distribution function}
There still remains an obstacle for numerical calculation.
If $v_\lambda(\br-\br')$ has infinitely strong repulsion,
Eqs.~\eqref{eq:flow_Fn} and \eqref{eq:flow_rhom}
seem to have divergent parts.
However, this seeming divergence can be eliminated
by introducing
the cavity distribution function \cite{han13}.
The $m$-particle cavity distribution function \cite{mee68} is defined by
\begin{align}
	\label{eq:def_y}
	y^{(m)}_\lambda(\br_1,\ldots,\br_m)
	=
	e^{\beta \sum_{i<j}^{m} v_\lambda (\br_i-\br_j)}
	g^{(m)}_\lambda(\br_1,\ldots,\br_m),
\end{align}
where 
$g^{(m)}_\lambda(\br_1,\cdots,\br_m)$ is 
the $m$-particle distribution functions,
which are defined by the normalization of
$\rho^{(m)}_\lambda(\br_1,\ldots,\br_m)$
with densities:
\begin{align}
	\label{eq:def_g}
	g^{(m)}_\lambda(\br_1,\ldots,\br_m)
	=
	\frac{\rho^{(m)}_\lambda(\br_1,\ldots,\br_m)}{\Pi_{i=1}^{m}\rho(\br_i)}.
\end{align}
The physical meaning of the cavity distribution function
is revealed by remembering Eqs.~\eqref{eq:def_ave} and \eqref{eq:rn_def},
which give
\begin{align}
	\label{eq:rhom_int}
	\rho^{(m)}_{\lambda}(\br_1,\ldots,\br_m)
	=&
	\frac{1}{\Xi_\lambda[\bp]}
	\sum_{N=m}^{\infty}
	\frac{1}{\Lambda^{3N} (N-m)!}
	\int_{\br_{m+1},\cdots,\br_N}
	\notag
	\\
	&\times
	e^{
	-\sum_{k<l}^{N}\beta v_\lambda(\br_{k}-\br_{l})
	+\sum_{k=1}^{N}\bp(\br_{k})
	}.
\end{align}
Rewriting Eq.~\eqref{eq:def_y} using Eqs.~\eqref{eq:def_g} and \eqref{eq:rhom_int},
one finds that the factor $\exp(-\sum_{k<l}^m \beta v_\lambda(\br_k-\br_l))$
coming from the interaction among the $m$ particles is canceled by
the exponential factor in Eq.~\eqref{eq:def_y}.
Therefore, $y^{(m)}_\lambda(\br_1,\ldots,\br_m)$ is proportional to
the probability that the $m$ particles which do not interact with each other
but interact with other particles are at $\br_1,\ldots,\br_m$.
Particularly in the case of hard spheres with a diameter $d$, 
the $m$ particles play an equivalent role to that of cavities with radius at least $d$
\cite{mee68}.
Since the $m$ particles can overlap with each other,
$y^{(m)}_\lambda(\br_1,\ldots,\br_m)$ can take a finite value
for $|\br_i-\br_j|<d$ with $i\neq j$ 
even in the hard-sphere case in contrast to $g^{(m)}_\lambda(\br_1,\ldots,\br_m)$.

Actually, the flow equations are rewritten as follows:
\begin{widetext}
\begin{align}
	\label{eq:flow_helm_inhom}
	&\partial_\lambda
	\beta F_\lambda
	=
	-
	\frac{1}{2}
	\int_{\br,{\bf r'}}
	\rho(\br)\rho(\br')
	\left[
	\partial_\lambda
	e^{-\beta v_\lambda(\br-\br')}
	\right]
	y^{(2)}_\lambda(\br,\br'),
	\\
	\label{eq:flow_y_inhom}
	&\partial_\lambda y_\lambda^{(m)}(\br_1,\ldots,\br_m)
	\notag
	\\
	&\qquad
	=
	\int_{\br}
	\rho(\br)
	\left[
	e^{-\beta \sum_{i}^{m}v_\lambda(\br_i-\br)}
	y^{(m+1)}_\lambda(\br_1,\ldots,\br_m,\br)
	-
	y^{(m)}_\lambda(\br_1,\ldots,\br_m)
	\right]
	\partial_\lambda \bp_{\max,\lambda}(\br)
	\notag
	\\
	&\qquad
	+
	y^{(m)}_\lambda(\br_1,\ldots,\br_m)
	\sum_{i=1}^m
	\partial_\lambda \bp_{\max,\lambda}(\br_i)
	+
	\int_{\br}
	\rho(\br)
	\left[
	\partial_\lambda
	e^{-\sum_{i}^{m}\beta v_\lambda(\br_i-\br)}
	\right]
	y^{(m+1)}_\lambda(\br,\br_1,\ldots,\br_m)
	\notag
	\\
	&\qquad
	+
	\frac{1}{2}
	\int_{\br,\br'}
	\rho(\br)\rho(\br')
	\left[
	\partial_\lambda
	e^{-\beta v_\lambda(\br-\br')}
	\right]
	\notag
	\\
	&\qquad
	\times
	\left[
	e^{
	-\sum_{i}^{m}\beta v_\lambda(\br_i-\br)
	-\sum_{i}^{m}\beta v_\lambda(\br_i-\br')}
	y^{(m+2)}_\lambda(\br,\br',\br_1,\ldots,\br_m)
	-
	y^{(2)}_\lambda(\br,\br')
	y^{(m)}_\lambda(\br_1,\ldots,\br_m)
	\right],
\end{align}
\end{widetext}
in which $v_\lambda (\br-\br')$ always appears
in the form of $e^{-\beta v_\lambda (\br-\br')}$
and there is no divergence due to the strong repulsion.
Therefore, these equations enable one 
to treat the short-range strong repulsion 
as well as the long-range attraction on the same footing.
In passing, each term in the right-hand side of Eq.~\eqref{eq:flow_y_inhom}
is interpreted as follows:
The first and second terms are derived from the change of $\bp_{\max,\lambda}(\br)$,
i.e., the change of the chemical potential to fix the density.
The third and fourth terms reflect
the changes of the interactions between any one of particles on $\br_1,\ldots,\br_m$
and another one
and between two particles other than particles on $\br_1,\ldots,\br_m$, respectively.
Since the cavity distribution function $y_\lambda^{(m)}(\br_1,\ldots,\br_m)$
is interpreted as the distribution 
when the interactions among particles on $\br_1,\ldots,\br_m$ are switched off \cite{han13}, 
there is no term reflecting the changes of interactions among them,
such as the second term in the left-hand side of Eq.~\eqref{eq:flow_rhom}.

\subsection{Truncation}

Although Eqs.~\eqref{eq:flow_helm_inhom}
and \eqref{eq:flow_y_inhom} are exact,
they depend on higher-order distribution functions
to form infinite hierarchy of coupled equations
and truncation at some order is needed for practical use.
Some approximate relation connecting distribution functions
of different order is required to close the hierarchy of equations.
Among many approximations proposed for $g^{(m)}_\lambda(\br_1,\ldots,\br_m)$,
preferable ones for our formulation
are those guaranteeing the finiteness of $y^{(m)}_\lambda(\br_1,\ldots,\br_m)$
in the case of strong repulsion.
A naive and suitable one is 
KSA, 
which exactly holds at low-density limit.
For simplicity, we consider KSA for
$y^{(3)}_\lambda(\br_1,\br_2,\br_3)$ and
$y^{(4)}_\lambda(\br_1,\br_2,\br_3,\br_4)$,
on which the second order flow equation depends:
\begin{widetext}
\begin{align}
	\label{eq:ksa3}
	y^{(3)}_\lambda (\br_1,\br_2,\br_3)
	&\approx
	y^{(2)}_\lambda(\br_1,\br_2)
	y^{(2)}_\lambda(\br_2,\br_3)
	y^{(2)}_\lambda(\br_3,\br_1)
	+
	\mathcal{O}(\rho),
	\\
	\label{eq:ksa4}
	y^{(4)}_\lambda (\br_1,\br_2,\br_3,\br_4)
	&\approx
	\frac{
	y^{(3)}_\lambda(\br_1,\br_2, \br_3)
	y^{(3)}_\lambda(\br_2,\br_3, \br_4)
	y^{(3)}_\lambda(\br_3,\br_4, \br_1)
	y^{(3)}_\lambda(\br_4,\br_1, \br_2)
	}
	{
	y^{(2)}_\lambda(\br_1,\br_2)	
	y^{(2)}_\lambda(\br_1,\br_3)	
	y^{(2)}_\lambda(\br_1,\br_4)	
	y^{(2)}_\lambda(\br_2,\br_3)	
	y^{(2)}_\lambda(\br_2,\br_4)	
	y^{(2)}_\lambda(\br_3,\br_4)	
	}
	+
	\mathcal{O}(\rhoi)
	\notag
	\\
	&\approx
	y^{(2)}_\lambda(\br_1,\br_2)
	y^{(2)}_\lambda(\br_1,\br_3)
	y^{(2)}_\lambda(\br_1,\br_4)
	y^{(2)}_\lambda(\br_2,\br_3)
	y^{(2)}_\lambda(\br_2,\br_4)
	y^{(2)}_\lambda(\br_3,\br_4)
	+
	\mathcal{O}(\rhoi).
\end{align}
\end{widetext}
A naive way to improve the accuracy of the calculation is
taking higher-order flow equations into account.
However, it may be time consuming to solve higher-order flow equations
since the number of arguments for the distribution functions increases.
Another approach for the improvement of the accuracy
is to find more accurate approximation beyond KSA.
There are many studies for the correction to KSA \cite{gro04}.
For example, according to Ref.~\cite{abe59}, the correction term 
to $y^{(3)}_\lambda(\br_1,\br_2, \br_3)$ 
up to $\mathcal{O}(\rhoi)$ is given by
\begin{widetext}
\begin{align}
	y^{(3)}_\lambda(\br_1,\br_2,\br_3)
	\approx
	&
	y^{(2)}_\lambda(\br_1,\br_2)
	y^{(2)}_\lambda(\br_2,\br_3)
	y^{(2)}_\lambda(\br_3,\br_1)
	e^{\rho\int_{\br}
	h_\lambda^{(2)}(\br_1,\br)
	h_\lambda^{(2)}(\br_2,\br)
	h_\lambda^{(2)}(\br_3,\br)}
	+
	\mathcal{O}(\rhoi^2),
\end{align}
\end{widetext}
where $h_\lambda^{(2)}(\br,\br')=g_\lambda^{(2)}(\br,\br')-1$
is the total correlation function,
in which $g_\lambda^{(2)}(\br,\br')$ is related 
to $y_\lambda^{(2)}(\br,\br')$ through Eq.~\eqref{eq:def_y}.
The convolution approximation \cite{ich70} is another approximation for
$g^{(3)}_\lambda(\br_1,\br_2,\br_3)$.
Unfortunately, this approximation does not guarantee
the finiteness of $y^{(3)}_\lambda(\br_1,\br_2,\br_3)$
since the core condition, which means that
$g^{(3)}_\lambda(\br_1,\br_2,\br_3)$ vanishes
if any of $|\br_i-\br_j|$ ($1\leq i < j \leq 3$) is lesser than
the hard-core diameter for hard-core systems, is not satisfied.

\subsection{Homogeneous cases}

In Sec.~\ref{sec:demo}, we will apply the formalism to homogeneous liquids.
For this purpose, we rewrite the flow equations for the homogeneous case.
In this case, $\rhoi(\br)$ and $\bp_{\max,\lambda}(\br)$ do not depend on $\br$
and $y^{(m)}_\lambda(\br_1,\ldots,\br_m)$ only depends 
on $\br_1-\br_m,\ldots,\br_{m-1}-\br_m$;
we denote $\rhoi(\br)$ and $\bp_{\max,\lambda}(\br)$
as $\rhoi$ and $\bp_{\max,\lambda}$, respectively,
and $y^{(m)}_\lambda(\br_1,\ldots,\br_m)$
as $y^{(m)}_\lambda(\br_1,\ldots,\br_{m-1})$
regarding $\br_m$ as the original point.
Then, Eqs.~\eqref{eq:flow_helm_inhom} and \eqref{eq:flow_y_inhom}
are respectively reduced to
\begin{widetext}
\begin{align}
	\label{eq:flow_helm_hom}
	&\partial_\lambda
	\frac{\beta F_\lambda}{N}
	=
	-
	\frac{\rhoi}{2}
	\int_{\br}
	\left[
	\partial_\lambda
	e^{-\beta v_\lambda(\br)}
	\right]
	y^{(2)}_\lambda(\br),
	\\
	\label{eq:flow_y_hom}
	&
	\partial_\lambda \ln y_\lambda^{(m)}(\br_1,\ldots,\br_{m-1})
	\notag
	\\
	&=
	\partial_\lambda \overline{\psi}_{{\max},\lambda}
	\left(
	m
	+
	\rhoi
	\int_{\br}
	\left[
	e^{-\beta \sum_{i}^{m-1}v_\lambda(\br_i-\br)
	-\beta v_\lambda(\br)}
	\frac{y^{(m+1)}_\lambda(\br_1,\ldots,\br_{m-1},\br)}
	{y^{(m)}_\lambda(\br_1,\ldots,\br_{m-1})}
	-
	1
	\right]
	\right)
	\notag
	\\
	&
	\quad
	+
	\rhoi
	\int_{\br}
	\partial_\lambda
	\left[e^{
	-\beta \sum_{i}^{m-1}v_\lambda(\br_i-\br)
	-\beta v_\lambda(\br)
	}\right]
	\frac{y^{(m+1)}_\lambda(\br,\br_1,\ldots,\br_{m-1})}
	{y^{(m)}_\lambda(\br_1,\ldots,\br_{m-1})}
	\notag
	\\
	&
	\quad
	+
	\frac{\rhoi^2}{2}
	\int_{\br,\br'}
	\partial_\lambda\left[e^{-\beta v_\lambda(\br-\br')}\right]
	y^{(2)}_\lambda(\br-\br')
	\notag
	\\
	&
	\quad
	\times
	\left[
	e^{
	-\beta \sum_{i}^{m-1}v_\lambda(\br_i-\br)
	-\beta \sum_{i}^{m-1}v_\lambda(\br_i-\br')
	-\beta v_\lambda(\br)
	-\beta v_\lambda(\br')
	}
	\frac{
	y^{(m+2)}_\lambda(\br,\br',\br_1,\ldots,\br_{m-1})
	}
	{y^{(2)}_\lambda(\br-\br')
	y^{(m)}_\lambda(\br_1,\ldots,\br_{m-1})}
	-
	1
	\right],
\end{align}
where $N=\rhoi\int d\br$ is the total particle number.
In this case, the flow equation for $\overline{\psi}_{{\max},\lambda}$
is explicitly derived from Eq.~\eqref{eq:flow_y_hom} for $m=1$ and
$y^{(1)}_\lambda=1$:
\begin{align}
	\label{eq:flow_psi}
	&\partial_\lambda \overline{\psi}_{{\max},\lambda}
	=
	-
	\frac{
	\rhoi
	\int_{\br}
	\partial_\lambda
	\left[e^{-\beta v_\lambda(\br)}\right]
	y^{(2)}_\lambda(\br)
	\left(
	1
	+
	\frac{\rhoi}{2}
	\int_{\br'}
	\left[
	e^{
	-\beta v_\lambda(\br+\br')
	-\beta v_\lambda(\br')
	}
	\frac{
	y^{(3)}_\lambda(\br+\br',\br')
	}{y^{(2)}_\lambda(\br)}
	-
	1
	\right]
	\right)
	}
	{
	1
	+
	\rhoi
	\int_{\br}
	\left[
	e^{-\beta v_\lambda(\br)}y^{(2)}_\lambda(\br)
	-
	1
	\right]
	}.
\end{align}
\end{widetext}

In homogeneous systems, the leading order of the density expansion for 
$\omega_{\lambda}(\br_1)=\ln y_{\lambda}(\br_1)$ is known as \cite{han13}
\begin{align}
	\label{eq:de_lead}
	\omega_\lambda(\br_1)
	=
	\rhoi\int_{\br}f_\lambda(\br_1-\br)f_\lambda(\br)
\end{align}
with the Mayer function $f_\lambda(\br)=e^{-\beta v_\lambda(\br)}-1$.
This can be reproduced
from Eqs.~\eqref{eq:flow_psi} and \eqref{eq:flow_y_hom} for $m=2$.
If we ignore
the $\lambda$ dependence in the right-hand sides
of Eqs.~\eqref{eq:flow_psi} and \eqref{eq:flow_y_hom} for $m=2$
except for 
$\partial_\lambda \left[e^{-\beta v_\lambda(\br)}\right]$
and 
$\partial_\lambda\left[e^{-\beta v_\lambda(\br-\br')}\right]$,
Eqs.~\eqref{eq:flow_psi} and \eqref{eq:flow_y_hom} for $m=2$
are approximated as follows:
\begin{align}
	\label{eq:flow_psi_de}
	\partial_\lambda \overline{\psi}_{{\max},\lambda}
	&
	\approx
	-
	\rhoi
	\int_{\br}
	\partial_\lambda
	\left[e^{-\beta v_\lambda(\br)}\right],
	\\
	\label{eq:flow_y2_de}
	\partial_\lambda \omega_\lambda(\br_1)
	&
	\approx
	2\partial_\lambda \overline{\psi}_{{\max},\lambda}
	+
	\rhoi
	\int_{\br}
	\partial_\lambda
	\left[e^{-\beta v_\lambda(\br_1-\br)-\beta v_\lambda(\br)}\right].
\end{align}
Inserting Eq.~\eqref{eq:flow_psi_de} into Eq.~\eqref{eq:flow_y2_de} 
and integrating Eq.~\eqref{eq:flow_y2_de} with respect to $\lambda$,
we have Eq.~\eqref{eq:de_lead}.
By considering the $\lambda$ dependence 
of other factors
in addition to $\partial_\lambda \left[e^{-\beta v_\lambda(\br)}\right]$ and 
$\partial_\lambda\left[e^{-\beta v_\lambda(\br-\br')}\right]$,
the higher-order 
contributions in the density expansion
are incorporated in our flow equations.

\subsection{Setting of the evolution to treat hard core\label{sec:evolution_hc}}
In principle, results do not depend on the choice of $v_{\lambda}(\br-\br')$
during the flow if the flow equations are exactly treated.
The introduction of truncation, however, causes the dependence
and one should discuss appropriate settings for $v_{\lambda}(\br-\br')$.
Moreover, some choice of $v_{\lambda}(\br-\br')$ 
causes breakdown of the numerical calculation.
Let us give a discussion about the choice of $v_{\lambda}(\br-\br')$
for a two-body interaction having a repulsive core in the case of employing KSA.

A naive choice may be the adiabatic-connection-like form:
\begin{align}
	\label{eq:ad}
	v_{\lambda}(\br-\br')
	=
	\lambda v(\br-\br').
\end{align}
This choice is, however, problematic in the case of the presence of a repulsive core.
For simplicity, let us consider the case of a hard sphere with a diameter $\sigma$.
Under the choice of Eq.~\eqref{eq:ad}, we have
\begin{align}
	e^{-\beta v_{\lambda}(\br-\br')}
	=
	\begin{cases}
		1 & \lambda=0,
		\\
		\theta(|\br-\br'|-\sigma) & \lambda > 0,
	\end{cases}
\end{align}
which shows the factors such as
$e^{-\beta v_{\lambda}(\br-\br')}$
appearing in the flow equations \eqref{eq:flow_helm_hom} and \eqref{eq:flow_y_hom}
suddenly changes as $\lambda$ departs from $0$
and particularly the divergences of the factors
$\partial_\lambda e^{-\beta v_{\lambda}(\br-\br')}$ occur at $\lambda=0$.
Such divergences are hard to treat in the numerical calculation.
\begin{figure}[!tb]
  \begin{center}
    \includegraphics[width=\columnwidth]{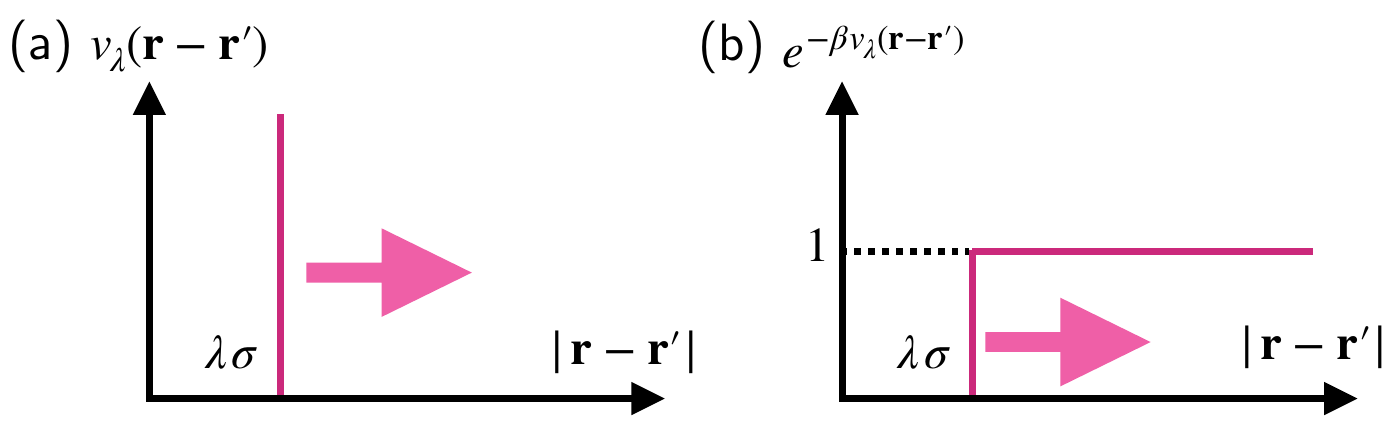}
    \caption{Visual representation of 
    (a) $v_{\lambda}(\br-\br')$ and (b) $e^{-\beta v_{\lambda}(\br-\br')}$
    under the evolution given by Eq.~\eqref{eq:rgflow}.}
    \label{fig:hs}
  \end{center}
\end{figure}

Our proposal for a possible choice not only circumventing this problem but also
contributing to the improvement of the accuracy with KSA
is the following one:
\begin{align}
	\label{eq:rgflow}
	v_{\lambda}(\br-\br')
	=
	\begin{cases}
		\infty & |\br-\br'|< \lambda\sigma,
		\\
		0 & |\br-\br'| \geq \lambda\sigma,
	\end{cases}
\end{align}
which is represented visually in Fig.~\ref{fig:hs}.
This evolution represents that the repulsive core is gradually
taken from short-range region and, in this sense,
is regarded as a choice inspired by the notion of RG.
Hereafter, we call the choice the RG-inspired flow.
In this case, the exponential factor and the derivative
become
\begin{align}
	e^{-\beta v_{\lambda}(\br-\br')}
	=&
	\theta(|\br-\br'|-\lambda\sigma),
	\\
	\partial_\lambda
	e^{-\beta v_{\lambda}(\br-\br')}
	=&
	-\sigma \delta(|\br-\br'|-\lambda\sigma).
\end{align}
Figure \ref{fig:hs} also shows the evolution of $e^{-\beta v_{\lambda}(\br-\br')}$.
Although a delta function appears in 
$\partial_\lambda e^{-\beta v_{\lambda}(\br-\br')}$,
this factor always appears in the integrands of spatial integrals
in Eqs.~\eqref{eq:flow_helm_hom} and \eqref{eq:flow_y_hom}
and does not cause divergence in the flow equations.
This choice also has an advantage for the accuracy of the calculation with KSA.
Since the only dimensionless parameter of the hard-sphere fluid 
is the packing fraction, 
a system composed of hard spheres with a small diameter
can be viewed as a low-density system.
In addition, the RG-inspired flow is equivalent
to considering an evolving packing fraction 
$\eta_\lambda=\rhoi V_{\rm HS}(\lambda\sigma)$
with $V_{\rm HS}(\lambda\sigma)$ being the volume of a hard sphere 
having a diameter $\lambda\sigma$.
Therefore, it is guaranteed that the flow equations with KSA becomes accurate
at least small $\lambda$ since KSA is accurate in low-density cases.
Although KSA holds only for low-density cases,
the property that the flow is accurate at least 
near the starting point $\lambda=0$ may contribute
to the improvement of the accuracy of FRG
in moderate and high densities.

It is conceivable to improve the accuracy by optimizing the evolution with a help of the principle of minimal sensitivity (PMS) \cite{can03}. In the PMS, one uses the exact condition by which the results do not depend on the choice of the evolution, which in our case, corresponds to generalizing the evolution of $\beta v_\lambda(\br-\br')$ for finding a stationary path.

The derivation of the flow equations for cavity distribution functions
and the ideas for truncation and choice of the flow
have been presented in this section.
In the next section, the accuracy of our method
will be assessed through 
applications to exactly solvable models.

\section{Demonstration in one-dimensional liquids\label{sec:demo}}
In this section, we apply the flow equations obtained in the previous section
to one-dimensional exactly solvable models to investigate the accuracy.
In particular, our results are compared to those obtained by HNC and PY.

\subsection{Model}
For the purpose of investigating the accuracy of our method, 
an application to a model for which exact solutions are obtained is desirable.
We employ a one-dimensional fluid composed of 
hard rods with an attractive force \cite{arc17}, whose interaction is described by
\begin{align}
	\beta v(r)
	=
	\begin{cases}
		\infty & |r|<\sigma
		\\
		-z_{\rm p}(\sigma+\sigma_{\rm p}-|r|) & \sigma \leq |r|<\sigma+\sigma_{\rm p}
		\\
		0 & \sigma+\sigma_{\rm p} \leq |r|
	\end{cases},
\end{align}
where $\sigma \geq 0$ is the diameter of a hard rod and
the parameters for the attractive force satisfies
$z_{\rm p} \geq 0$ and $\sigma_{\rm p} \geq 0$.

It is known that exact solutions for
the structure factor and thermodynamic quantities
are obtained in the case of $\sigma_{\rm p}<\sigma$,
which implies that each particle interacts with only the nearest neighbor ones
and simplifies the calculation of the grand partition function.
The pressure $P$ is obtained by solving the following equation \cite{bra02}:
\begin{align}
	\frac{1}{\rhoi}=-\left.\frac{d}{ds}\ln\tilde{J}(s)\right|_{s=\beta P},
\end{align}
where $\tilde{J}(s)$ is defined by the following Laplace transform:
\begin{align}
	\tilde{J}(s)=\int_0^{\infty}dx e^{-sx}e^{-\beta v(x)}.
\end{align}
The chemical potential is determined as a function of $\beta P$:
\begin{align}
	\beta \mu(\beta P)=\ln\frac{\Lambda}{\tilde{J}(\beta P)}.
\end{align}
Once the pressure is calculated, the structure factor
is obtained through the exact relation \cite{per82}:
\begin{align}
	S(k)=\frac{1-|a(k)|^2}{|1-a(k)|^2}
\end{align}
with $a(k)=e^{-\beta\mu(\beta P + ik)+\beta\mu(\beta P)}$.

\subsection{Details of calculation}
\begin{figure}[!tb]
  \begin{center}
    \includegraphics[width=\columnwidth]{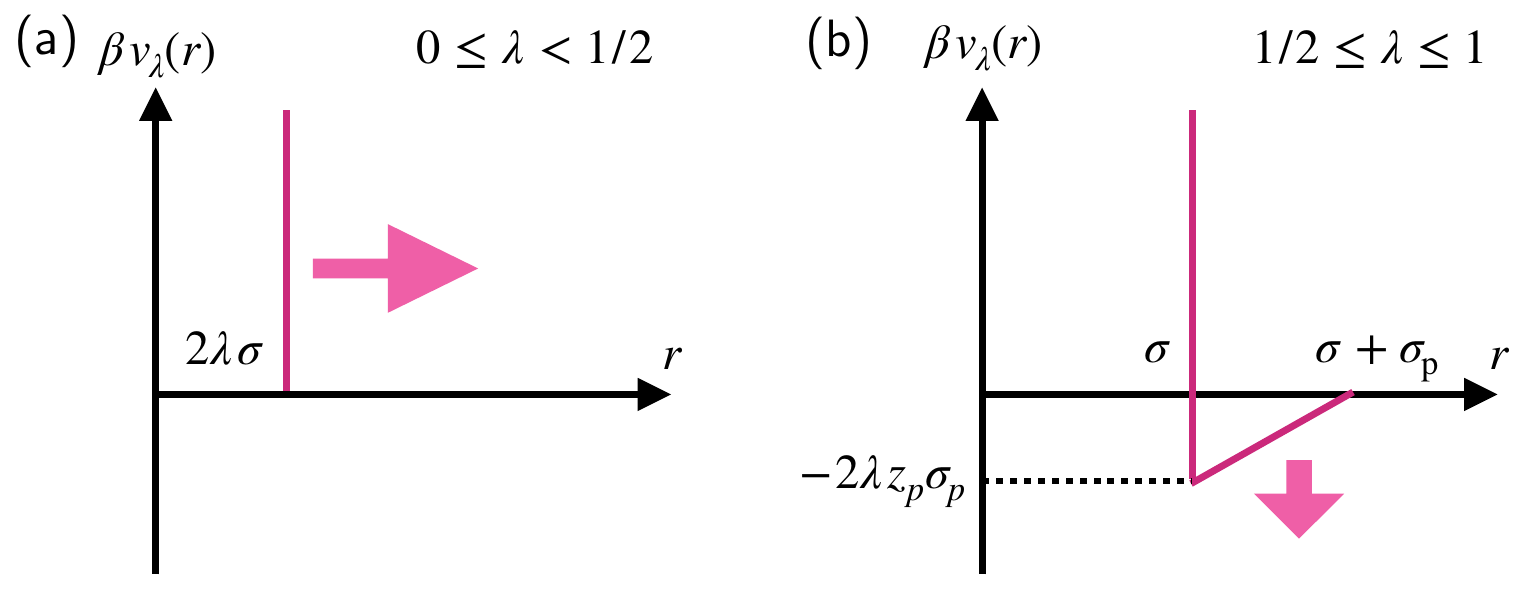}
    \caption{Visual representation of the evolution of $\beta v_{\lambda}(r)$
    in 
    (a) $0\leq \lambda< 1/2$ given by Eq.~\eqref{eq:evo_hs} and 
    (b) $1/2\leq \lambda\leq 1$ given by Eq.~\eqref{eq:evo_att}.}
    \label{fig:v_evo}
  \end{center}
\end{figure}
\subsubsection{FRG}
We consider the flow equations up to the second order,
i.e., Eqs.~\eqref{eq:flow_helm_hom}, \eqref{eq:flow_psi},
and \eqref{eq:flow_y_hom} for $m=2$.
KSA (Eqs.~\eqref{eq:ksa3} and  \eqref{eq:ksa4})
is applied to the higher-order correlation functions 
$y^{(3)}_\lambda(r_1,r_2)$ and $y^{(4)}_\lambda(r_1,r_2,r_3)$
appearing in the flow equations.
We set the evolution in two steps:
In the first step, the hard repulsive part of the interaction
is taken with the RG-inspired flow represented in Eq.~\eqref{eq:rgflow}.
We find that the application of the RG-inspired flow 
to the attractive
part sometimes makes the numerical calculation unstable,
which may be because the exponential factor
$e^{-\beta v_{\lambda}(r)}$ can take a large value 
when $\beta v_{\lambda}(r)<0$
and the evolution becomes unexpectedly rapid during the RG-inspired flow. 
Instead of this, as the second step, we incorporate the attractive part
with the adiabatic-connection-like flow.
In summary, $\beta v_{\lambda}(r)$ is set as follows:
\begin{align}
	\label{eq:evo_hs}
	\beta v_{\lambda}(r)
	=
	\begin{cases}
		\infty & |r|< 2\lambda\sigma
		\\
		0 & |r| \geq 2\lambda\sigma
	\end{cases},
\end{align}
for $0\leq \lambda < 1/2$, and
\begin{align}
	\label{eq:evo_att}
	\beta v_\lambda(r)
	=
	\begin{cases}
		\infty & |r|<\sigma
		\\
		-(2\lambda-1)z_{\rm p}(\sigma+\sigma_{\rm p}-|r|) & \sigma \leq |r|<\sigma+\sigma_{\rm p}
		\\
		0 & \sigma+\sigma_{\rm p} \leq |r|
	\end{cases},
\end{align}
for $1/2\leq \lambda \leq 1$.
This choice for $\beta v_{\lambda}(r)$ is visualized in Fig.~\ref{fig:v_evo}.
Since the interaction is turned off at $\lambda=0$,
the initial condition required for solving the flow equations
is given by $y^{(2)}_{\lambda=0}(r)=1$.

For the numerical implementation of our calculation,
the GNU scientific library (GSL) is employed.
In both the first and second parts of the evolution, i.e.~$0\leq \lambda < 1/2$
and $1/2\leq \lambda \leq 1$, respectively,
the flow equations are solved with the eighth-order Runge--Kutta method
within 20 steps.
The cavity distribution function $y^{(2)}_\lambda(r)$
is derived on 512 grid points in $0 \leq r<10\sigma$.
To evaluate the spatial integral, we employ
the Gauss--Krnord 21-point method with applying the spline interpolation 
for $y^{(2)}_\lambda(r)$.
\begin{figure}[!tb]
  \begin{center}
    \includegraphics[width=\columnwidth]{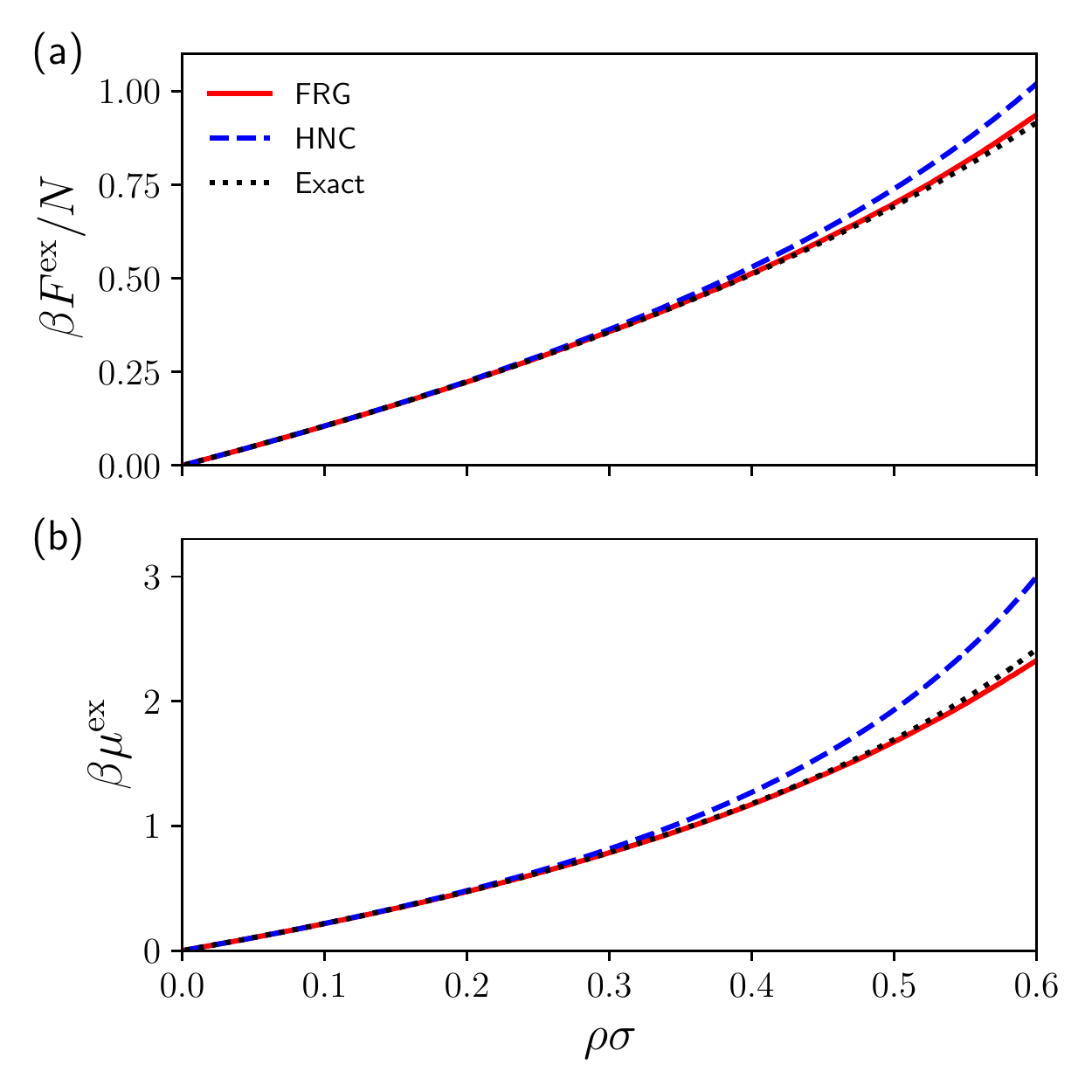}
    \caption{(a) Excess free energy per particle $\beta F^{\rm ex}/N$ and (b)  chemical potential $\beta \mu^{\rm ex}$ of pure hard rods
    as functions of the packing fraction $\rhoi\sigma$. The results given by FRG, HNC, and exact solution are depicted by the red solid, blue dashed, and black dotted lines, respectively. PY results are not explicitly shown since they are identical to the exact solutions.}
    \label{fig:thermo_hard}
  \end{center}
\end{figure}
\begin{figure*}[!htb]
  \begin{center}
    \includegraphics[width=2\columnwidth]{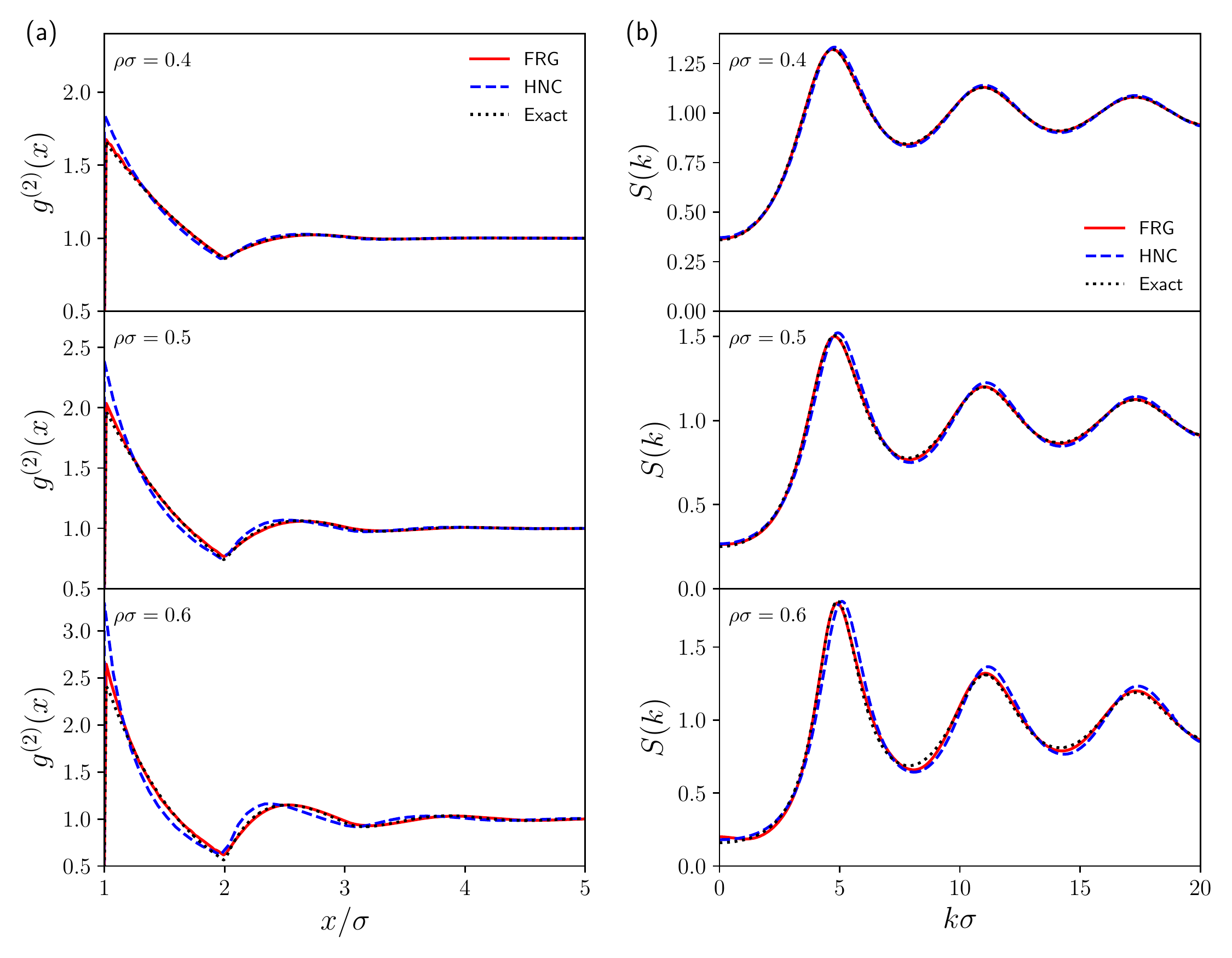}
    \caption{(a) Pair correlation functions $g^{(2)}(x)$ and (b) static structure functions $S(k)$ 
    of pure hard rods calculated in $\rhoi\sigma=0.4$, $0.5$, and $0.6$. The style and color of each line
    corresponding to each method is the same as Fig.~\ref{fig:thermo_hard}.}
    \label{fig:g_hard}
  \end{center}
\end{figure*}

\subsubsection{Integral-equation method}
For the purpose of comparing our method to other conventional ones,
we perform 
calculations by use of the integral-equation method with HNC and PY. 
In this method, the pair distribution function
is calculated through 
the coupled integral equations composed of
the OZ equation
\begin{align}
	h(r)=c(r)+\rhoi\int dr' c(r-r')h(r')
\end{align}
and a closure relation, which is given by
\begin{align}
	h(r)&=e^{-\beta v(r)+h(r)-c(r)}-1 &(\text{HNC}),
	\\
	\label{eq:py}
	h(r)&=e^{-\beta v(r)}\left[1+h(r)-c(r)\right]-1 &(\text{PY}).
\end{align}
Here, $c(r)$ is the direct correlation function and
$h(r)=g^{(2)}(r)-1$ is the total correlation function.
The coupled integral equations are solved numerically 
in an iterative manner. 
We use the exact solutions as the starting conditions of the iteration.
From the resultant $g^{(2)}(r)$,
we calculate $\beta P$ through the pressure equation:
\begin{align}
	\frac{\beta P}{\rhoi} = 1-\rhoi \int_0^{\infty} dr\, r\frac{d\beta v(r)}{dr}
	g^{(2)}(r).
\end{align}
In the case of HNC, the excess chemical potential $\mu^{\rm ex}$ can be calculated
using \cite{han13}
\begin{align}
	\beta \mu^{\rm ex}
	=
	\frac{1}{2}\rhoi\int dr
	h(r)\left[h(r)-c(r)\right]-\rhoi\int dr c(r).
\end{align}
In the case of PY, we calculate $\beta P(\rhoi)$ for various densities
and perform the numerical integration with respect to $\rhoi$ to obtain 
the free-energy per particle:
\begin{align}
	\frac{\beta F}{N}=\int_0^{\infty}d\rhoi'\frac{\beta P(\rhoi')}{\rhoi^{\prime 2}},
\end{align}
which is derived from the thermodynamic relation:
\begin{align}
	P =-\left(\frac{\partial F}{\partial V}\right)_{\beta,N},
\end{align}
with the spatial volume $V$.
The chemical potential is also obtained from 
\begin{align}
	\label{eq:mufp}
	\beta\mu=\frac{F}{N}+\frac{\beta P}{\rhoi}.
\end{align}

\subsection{Results for pure hard rods}
We first show the case of a liquid 
composed of pure hard rods (Tonks gas), i.e., the case of $z_{\rm p}=0$.
As for the thermodynamic quantities,
the excess free energy per particle
$F^{\rm ex}/N=F_{\lambda=1}/N-F_{\lambda=0}/N$ 
and the excess chemical potential
$\mu^{\rm ex}=(\bp_{\max,\lambda=1}-\bp_{\max,\lambda=0})/\beta$ 
are obtained from FRG.
Figure \ref{fig:thermo_hard} shows the results of $\beta F^{\rm ex}/N$
and $\beta \mu^{\rm ex}$
together with the exact solutions and those obtained by HNC
as functions of the packing fraction $\rho\sigma$.
For the pure hard-rod system, PY gives the exact solutions \cite{wer64, ver64}.
On the other hand, the results by HNC deviates from the exact solutions
as the system becomes dense.
In both results of $\beta F^{\rm ex}/N$ and $\beta \mu^{\rm ex}$,
FRG shows more accurate results than those given by HNC.

Figure \ref{fig:g_hard} shows the results of 
the pair distribution function $g^{(2)}(r)$
and the static structure factor $S(k)$ for some packing fractions.
Through the numerical evaluation of the Fourier transform
\begin{align}
	g^{(2)}(r)=\frac{1}{\rhoi}\int_{-\infty}^{\infty} \frac{dk}{2\pi}e^{ikr}\left[S(k)-1\right]+1,
\end{align}
$g^{(2)}(r)$ is calculated from $S(k)$ in the case of the exact solution
and vice versa in the cases of FRG and the integral-equation method.
HNC misses the height of the first peak and
the position of the second peak of $g^{(2)}(r)$,
which becomes worse as the system becomes dense.
In comparison to HNC, FRG gives more accurate results, 
although it slightly overestimates
the height of the first peak at high density.
FRG also gives more accurate results
for the height and position of each peak of $S(k)$ than HNC.
We also calculate the pressure
from $F^{\rm ex}/N$ and $\beta\mu^{\rm ex}$ via Eq.~\eqref{eq:mufp}.
The result of the pressure by FRG is accurate as well as $F^{\rm ex}/N$ and $\beta\mu^{\rm ex}$.

\subsection{Results for hard rod with attractive force}
\begin{figure}[!tb]
  \begin{center}
    \includegraphics[width=\columnwidth]{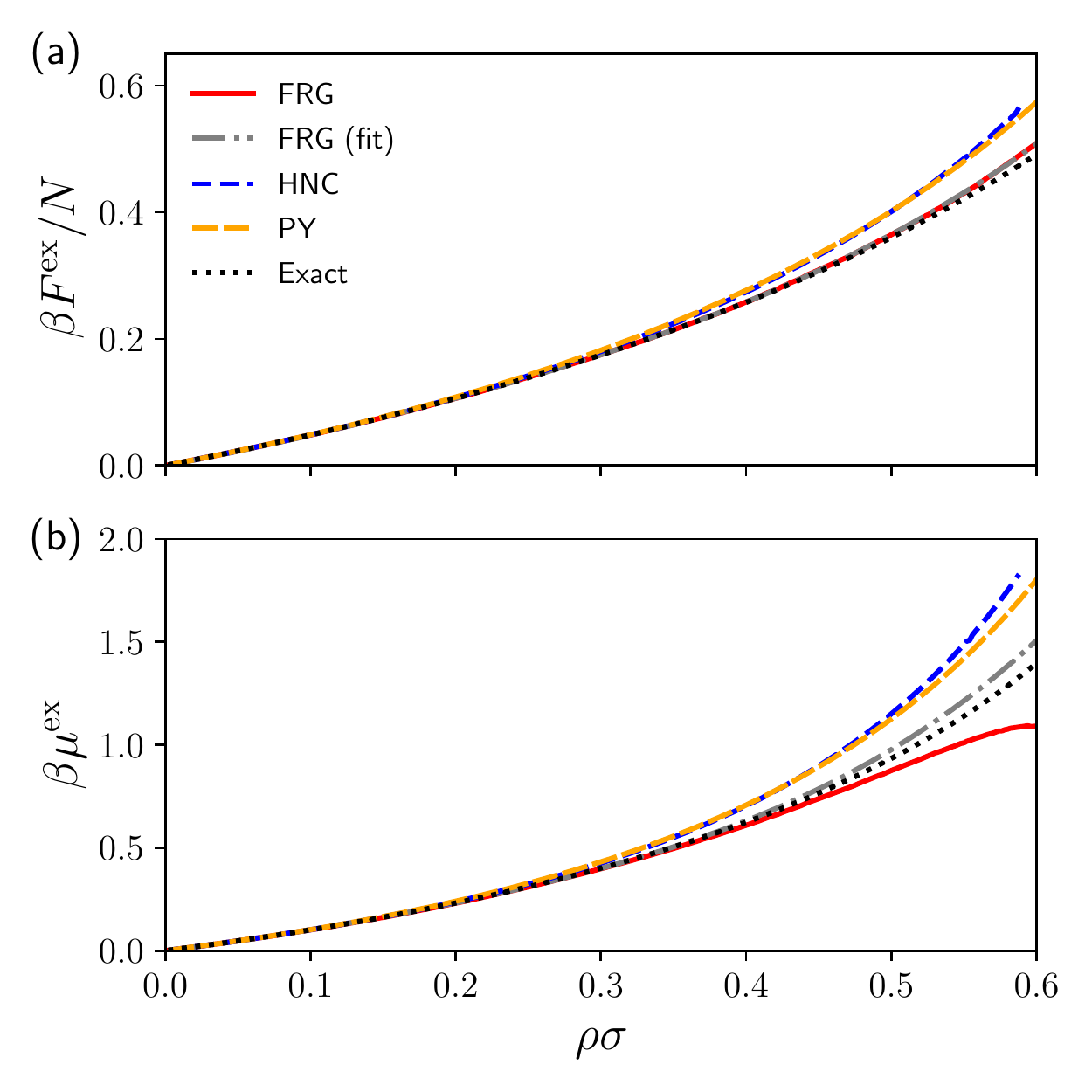}
    \caption{(a) Excess free energy per particle $\beta F^{\rm ex}/N$ 
    and (b) chemical potential $\beta \mu^{\rm ex}$ 
    in the case of $z_{\rm p}=1$ and $\sigma_{\rm s}=0.9$
    as functions of the packing fraction $\rhoi\sigma$. 
    The result of the fitting of $\beta F^{\rm ex}/N$ given by FRG
    with a fourth-order polynomial and
    the chemical potential calculated from the fitting function
    are shown as a gray dotted--dashed line.
    The PY results are shown as the orange long-dashed lines.
    For other methods, the style and color of each line
    corresponding to each method
    is the same as Fig.~\ref{fig:thermo_hard}.}
    \label{fig:thermo_att}
  \end{center}
\end{figure}
\begin{figure*}[!tb]
  \begin{center}
    \includegraphics[width=2\columnwidth]{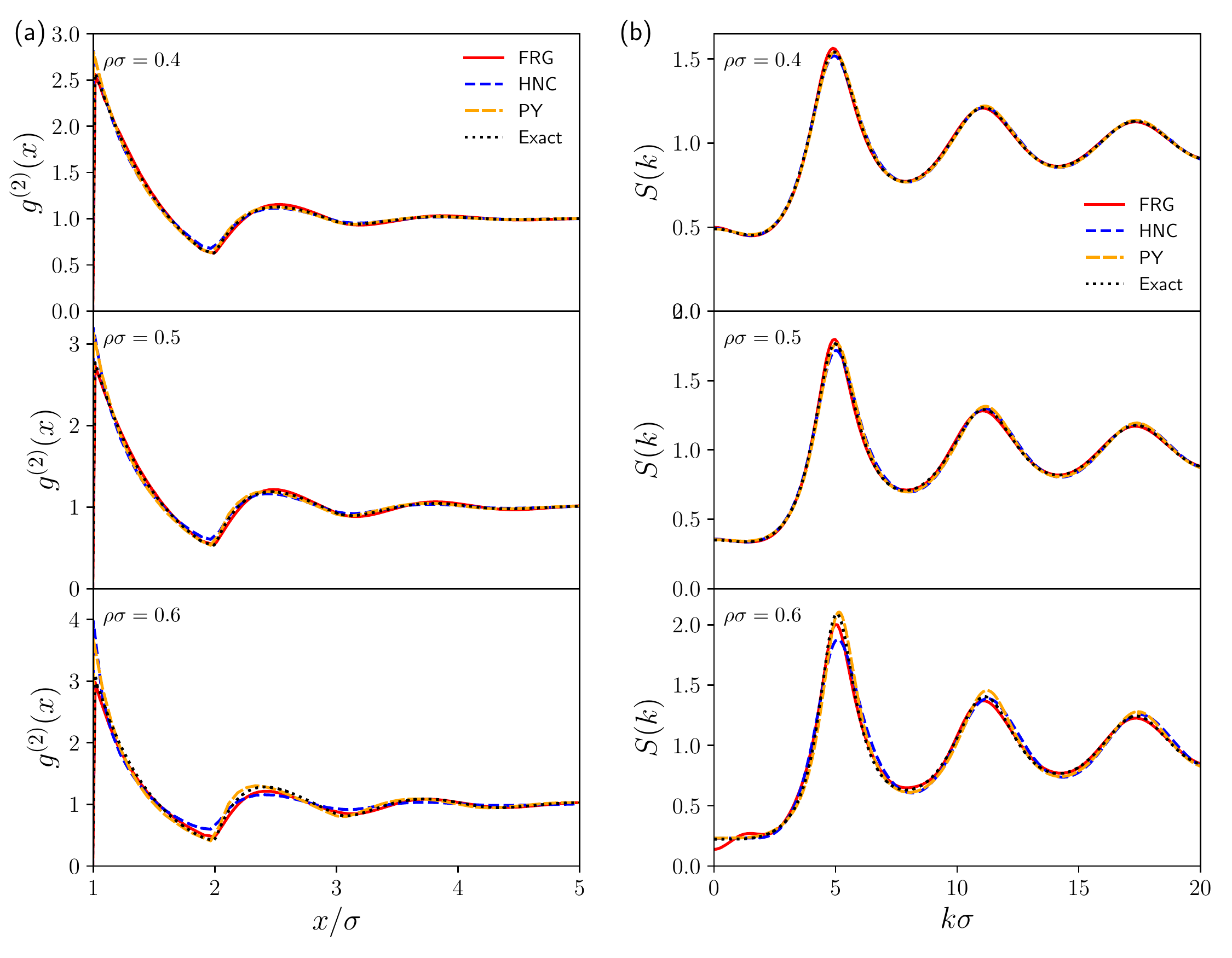}
    \caption{(a) Pair correlation functions $g^{(2)}(x)$ 
    and (b) static structure functions $S(k)$
    of hard rods with $z_{\rm p}=1$ and $\sigma_{\rm s}=0.9$    
    calculated in $\rhoi\sigma=0.4$, $0.5$, and $0.6$. The style and color of each line
    corresponding to each method is the same as Fig.~\ref{fig:thermo_att}.}
    \label{fig:g_att}
  \end{center}
\end{figure*}
We set $z_{p}=1$ and $\sigma_{p}=0.9$ to see the effect of the attractive force.
The results for the thermodynamic quantities are shown in Fig.~\ref{fig:thermo_att}.
The results of HNC is not shown in $\rhoi\sigma\geq 0.59$
since the calculation does not converge in this region.
Due to the presence of the attractive force, the results of PY no longer agree
with the exact solutions. It is noteworthy that 
FRG gives an accurate result for 
$\beta F^{\rm ex}/N$ in comparison to HNC and PY 
although PY has advantages for hard-rod systems
as it gives exact solutions without the attractive force.
On the other hand, an anomalous decrease is found near $\rhoi\sigma=0.6$
in the result of $\beta \mu^{\rm ex}$ by 
solving the FRG flow equation for the chemical potential \eqref{eq:flow_psi}.
The pressure obtained from Eq.~\eqref{eq:mufp} also shows such a qualitative failure.
The failure in contrast to the $\beta F^{\rm ex}/N$ may be because
the flow equation for $\bp_{\max,\lambda}$ is directly approximated with KSA
as shown in Eq.~\eqref{eq:flow_psi} while 
the flow equation for $\beta F_{\lambda}/N$
is affected by the approximation indirectly
through $y^{(2)}_\lambda(r)$ as shown in Eq.~\eqref{eq:flow_helm_hom}.
Actually, 
the accuracy for $\beta \mu^{\rm ex}$ can differ 
from that for $\beta F^{\rm ex}_{\lambda}/N$
since our approximation violates the thermodynamic relation between $\beta \mu^{\rm ex}$
and $\beta F^{\rm ex}_{\lambda}/N$:
\begin{align}
	\beta\mu^{\rm ex}
	=
	\left(
	\frac{\partial \beta F^{\rm ex}}{\partial N}
	\right)_{\beta,V}
	=
	\frac{\beta F^{\rm ex}}{N}
	+
	\rhoi
	\left(
	\frac{\partial}{\partial \rhoi}
	\frac{\beta F^{\rm ex}}{N}
	\right)_{\beta}.
\end{align}
Instead of the flow equation for $\bp_{\max,\lambda}$,
we also evaluate $\beta \mu^{\rm ex}$ from this thermodynamic relation
with evaluating the derivative of $\beta F^{\rm ex}_{\lambda}/N$ with respect to $\rhoi$.
We fit the result of $\beta F^{\rm ex}/N$
by a fourth-order polynomial $\beta F^{\rm ex}/N=\sum_{i=1}^{4}a_{i}(\rhoi\sigma)^i$
and calculate the derivative from the fitting function, 
instead of performing the numerical derivative, which gives noisy results.
The resultant fitting curve for $\beta F^{\rm ex}/N$
and $\beta \mu^{\rm ex}$ obtained by use of the fitting function
are depicted as a gray dotted--dashed lines in Fig.~\ref{fig:thermo_att}.
The result of $\beta \mu^{\rm ex}$ from the fitting-aided method
gives reasonable and more accurate result than other methods.

Finally, the results for $g^{(2)}(r)$ and $S(k)$ are presented.
Figure \ref{fig:g_att} shows the results of $g^{(2)}(r)$ and $S(k)$
calculated for some packing fractions.
Particularly, FRG accurately predicts the height of the first peak, while
other methods overestimate it as the system becomes dense.
For $S(k)$, FRG gives results comparably accurate to PY,
while there appears anomalous behavior near $k=0$ at high density.

\section{Conclusion\label{sec:con}}
In this paper, we present a method for classical liquids
based on the functional renormalization group (FRG).
The flow equations associated with
the evolution of the two-body interaction
are derived for the cavity distribution functions
at arbitrary order,
which are suitable to treat the short-range repulsion in the interaction.
As a practical prescription for numerical calculations,
we have proposed truncation method using the Kirkwood superposition approximation (KSA) and
pointed out that choosing the evolution
so that the interaction is gradually incorporated
from the short-range to long-range parts
is a suitable choice of the flow to treat a strong repulsive part with KSA.
To illustrate how our method works,
an application to a one-dimensional liquid 
composed of hard rods with and without an attractive force,
which is an exactly solvable model,
and the comparison to the integral-equation method based on the Ornstein-Zernike equation
such as the hypernetted chain (HNC) and the Percus-Yevick (PY) equation 
have been presented.
The flow equations up to the second order with KSA 
have been employed in the calculation.
It is noteworthy that the excess free energy per particle
and the height of the first peak of the pair distribution function $g^{(2)}(r)$
given by FRG
show better results than those given by PY in the presence of 
the attractive force 
although PY has advantages for hard rods as it gives exact solutions
without the attractive force. 
In comparison with HNC, FRG shows better results
for the excess free energy per particle
and the heights and positions of peaks 
in both cases with and without the attractive force.
These results suggest that FRG can become an accurate framework
to incorporate the repulsive and attractive parts of the interaction
on the same footing without a reference system
 representing the contribution from short-range repulsion.

To further improve the accuracy 
and avoid anomalous behaviors of the excess chemical potential
and structure functions in low momentum observed 
in our calculation in high density,
the improvement of the approximation is desirable.
In addition to treating higher-order flow equations,
we have referred to a way to taking correction terms
to KSA proposed in Ref.~\cite{abe59} into account.
Meanwhile, one may alternatively be based on small expansion parameters realized by using the gradient expansion, HRT, or the Blaizot--M\'endez--Wschebor approximation \cite{bla06} as discussed in Ref.~\cite{dup21}.
Although we have treated the short-range repulsion by introducing the cavity distribution functions, one may alternatively use the method developed for spin systems \cite{mac10}, where the short-range fluctuations are incorporated by solving single site problems and FRG is used to incorporate longer-range fluctuations.

Our method can be straightforwardly
extended to three-dimensional systems
and applied to other potentials such as the Lennard--Jones potential,
which are, of course, important future tasks.
Regarding the calculation in three-dimensional systems,
we believe that it is efficiently doable
with employing efficient ways to evaluate the spatial integrals \cite{bar62}.
The description of phase transition and the analysis of critical points,
to which previous studies with FRG for classical liquids are devoted, 
is another topic of interest.

\begin{acknowledgements}
	T.~Y.~was supported by the Grants-in-Aid for Japan Society for the Promotion of Science
	(JSPS) fellows (Grant No.~20J00644).
\end{acknowledgements}

\appendix
\begin{widetext}
\section{Derivation of Eqs.~\eqref{eq:flow_Fn} and \eqref{eq:flow_rhom}
\label{sec:derivation_flow}}
In this Appendix, we present the derivation of Eqs.~\eqref{eq:flow_Fn} and \eqref{eq:flow_rhom}.
For convenience, let us introduce the following notation
for $m$-particle density, in which the argument is explicitly shown:
\begin{align}
	\label{eq:rn_def_r}
	\rho^{(m)}_\lambda[\rhoa](\br_1,\ldots,\br_m)
	=&
	\left\langle
	\prod_{i=1}^{m}
	\left(
	\hat{\rho}(\br_i)
	-
	\sum_{j=1}^{i-1}\delta(\br_j-\br_i)
	\right)		
	\right\rangle_{\lambda,\bp_{\max,\lambda}[\rhoa]}.
\end{align}
From Eqs.~\eqref{eq:r2}, \eqref{eq:finv}, and \eqref{eq:rn_def_r}, we have
\begin{align}
	\label{eq:finv_r2}
	\left(
	\frac{\delta^2 \beta F_{\lambda}}{\delta \rhoa\delta \rhoa}
	\right)^{-1}[\rhoa](\br,\br')
	=&
	\left
	\langle
	\hat{\rho}(\br)
	\hat{\rho}(\br')
	\right
	\rangle_{\lambda,\bp_{\max,\lambda}[\rhoa]}
	-
	\rhoa(\br)\rhoa(\br')
	\notag
	\\
	=&
	\rho^{(2)}_\lambda[\rhoa](\br,\br')
	+
	\rhoa(\br)\delta(\br-\br')
	-
	\rhoa(\br)\rhoa(\br').
\end{align}
Inserting this into Eq.~\eqref{eq:Fflow_eq} with the substitution
$\rhoa(\br)=\rhoi(\br)$, we arrive at Eq.~\eqref{eq:flow_Fn}.

The derivation of Eq.~\eqref{eq:flow_rhom} is based on the mathematical induction.
In terms of $\rho^{(m)}_\lambda[\rhoa](\br_1,\ldots,\br_m)$,
Eq.~\eqref{eq:flow_rhom} corresponds to the following equation:
\begin{align}
	\label{eq:rhom_assume}
	&
	\partial_\lambda
	\rho_\lambda^{(k)}[\rhoa](\br_1,\ldots,\br_k)
	+
	\sum_{i<j}^{k}
	\partial_\lambda
	\left[\beta v_\lambda(\br_i-\br_j)\right]
	\rho_\lambda^{(k)}[\rhoa](\br_1,\ldots,\br_k)
	\notag
	\\
	&
	\quad
	=
	\int_{\br}
	\left[
	\rho^{(k+1)}_\lambda[\rhoa](\br_1,\ldots,\br_k,\br)
	-
	\rhoa(\br)
	\rho^{(k)}_\lambda[\rhoa](\br_1,\ldots,\br_k)
	\right]
	\partial_\lambda \bp_{\max,\lambda}[\rhoa](\br)
	\notag
	\\
	&
	\qquad
	+
	\rho^{(m)}_\lambda[\rhoa](\br_1,\ldots,\br_k)
	\sum_{i=1}^k
	\partial_\lambda \bp_{\max,\lambda}[\rhoa](\br_i)
	-
	\int_{\br}
	\sum_{i=1}^k
	\partial_\lambda
	\left[\beta v_\lambda(\br-\br_i)\right]
	\rho^{(k+1)}_\lambda[\rhoa](\br, \br_1, \ldots, \br_m)
	\notag	
	\\
	&
	\qquad
	-
	\frac{1}{2}
	\int_{\br,\br'}
	\partial_\lambda\left[\beta v_\lambda(\br-\br')\right]
	\left[
	\rho^{(k+2)}_\lambda[\rhoa](\br,\br',\br_1,\ldots,\br_m)
	-
	\rho^{(2)}_\lambda[\rhoa](\br,\br')
	\rho^{(k)}_\lambda[\rhoa](\br_1,\ldots,\br_k)
	\right].
\end{align}
Let us show this equation holds for arbitrary integer $k\geq 1$.

For the proof, we prepare an expression for the derivative of 
$\rho^{(m)}_\lambda[\rhoa](\br_1,\ldots,\br_m)$ with respect to $\rhoa(\br)$:
\begin{align}
	\label{eq:rhom_r_i}
	\frac{\delta \rho^{(m)}_\lambda[\rhoa](\br_1,\ldots,\br_m)}{\delta \rhoa(\br)}
	=&
	\left.
	\int_{\br_{m+1}}
	\frac{\delta \bp_{\max,\lambda}[\rhoa](\br_{m+1})}{\delta \rhoa(\br)}
	\frac{\delta \rho^{(m)}_\lambda[\rhoa](\br_1,\ldots,\br_m)}{\delta \bp(\br_{m+1})}
	\right|_{\bp=\bp_{\max,\lambda}[\rhoa]}
	\notag
	\\
	=&
	\left.
	\int_{\br_{m+1}}
	\frac{\delta^2 \beta F_\lambda[\rhoa]}{\delta \rhoa(\br)\delta \rhoa(\br_{m+1})}
	\frac{\delta \rho^{(m)}_\lambda[\rhoa](\br_1,\ldots,\br_m)}{\delta \bp(\br_{m+1})}
	\right|_{\bp=\bp_{\max,\lambda}[\rhoa]},
\end{align}
where we have used Eq.~\eqref{eq:F_psimax}.
Using Eq.~\eqref{eq:rn_def}
and remembering Eq.~\eqref{eq:def_ave}, we have
\begin{align}
	\label{eq:rn_dif}
	&\left.
	\frac{
	\delta\rho^{(m)}_\lambda[\rhoa](\br_1,\cdots,\br_m)
	}{\delta \bp(\br_{m+1})}
	\right|_{\bp=\bp_{\max,\lambda}[\rhoa]}
	\notag
	\\
	=&
	\left\langle
	\hat{\rho}({\bf r}_{m+1})
	\prod_{i=1}^{m}
	\left(
	\hat{\rho}({\bf r}_i)
	-
	\sum_{j=1}^{i-1}\delta({\bf r}_j-{\bf r}_i)
	\right)		
	\right\rangle_{\lambda,\bp_{\max,\lambda}[\rhoa]}
	-
	\rho({\bf r}_{m+1})
	\rho^{(m)}_\lambda[\rhoa](\br_1,\cdots,\br_m)
	\notag
	\\
	=&
	\rho^{(m+1)}_\lambda[\rhoa]({\bf r}_1,\cdots,{\bf r}_{m+1})
	+
	\rho^{(m)}_\lambda[\rhoa]({\bf r}_1,\cdots,{\bf r}_m)
	\left[
	\sum_{k=1}^{m}\delta({\bf r}_k-{\bf r}_{m+1})
	-
	\rhoa({\bf r}_{m+1})
	\right].
\end{align}
Therefore, Eq.~\eqref{eq:rhom_r_i} is rewritten as
\begin{align}
	\label{eq:rhom_r}
	&\frac{\delta \rho^{(m)}_\lambda[\rhoa](\br_1,\ldots,\br_m)}{\delta \rhoa(\br)}
	\notag
	\\
	=&
	\int_{\br'}
	\frac{\delta^2 \beta F_\lambda[\rhoa]}{\delta \rhoa(\br)\delta \rhoa(\br')}
	\left(
	\rho^{(m+1)}_\lambda[\rhoa]({\bf r}_1,\cdots,{\bf r}_{m},\br')
	+
	\rho^{(m)}_\lambda[\rhoa]({\bf r}_1,\cdots,{\bf r}_m)
	\left[
	\sum_{k=1}^{m}\delta({\bf r}_k-{\bf r}')
	-
	\rhoa({\bf r}')
	\right]	
	\right).
\end{align}

Equation \eqref{eq:rhom_assume} for $k=1$ is obtained from the first derivative
of Eq.~\eqref{eq:Fflow_eq}:
\begin{align}
	\label{eq:fflow_r1}
	\partial_\lambda
	\frac{\delta \beta F_\lambda[\rhoa]}{\delta \rhoa(\br_1)}
	=&
	\frac{1}{2}
	\int_{\br,\br'}
	\partial_\lambda
	\beta v_{\lambda}(\br-\br')
	\frac{\delta}{\delta\rhoa(\br_1)}
	\left(
	\rhoa(\br)\rhoa(\br')
	+
	\left(
	\frac{\delta^2 \beta F_{\lambda}}{\delta \rhoa\delta \rhoa}
	\right)^{-1}[\rhoa](\br,\br')
	-
	\rhoa(\br)\delta(\br-\br')
	\right).
\end{align}
From Eqs.~\eqref{eq:finv_r2} and \eqref{eq:rhom_r}, we obtain
\begin{align}
	\label{eq:finv_r2}
	&
	\frac{\delta}{\delta\rhoa(\br_1)}
	\left[
	\rhoa(\br)\rhoa(\br')
	+
	\left(
	\frac{\delta^2 \beta F_{\lambda}}{\delta \rhoa\delta \rhoa}
	\right)^{-1}[\rhoa](\br,\br')
	-
	\rhoa(\br)\delta(\br-\br')
	\right]
	\notag
	\\
	=&
	\int_{\br''}
	\frac{\delta^2 \beta F_\lambda[\rhoa]}{\delta \rhoa(\br_1)\delta \rhoa(\br'')}
	\left(
	\rho^{(3)}_\lambda[\rhoa](\br,\br',\br'')
	+
	\rho^{(2)}_\lambda[\rhoa](\br,\br')
	\left[
	\sum_{k=1}^{m}
	\delta(\br-\br'')
	+
	\delta(\br'-\br'')
	-
	\rhoa(\br'')
	\right]	
	\right).
\end{align}
By use of this relation and Eq.~\eqref{eq:F_psimax}, 
Eq.~\eqref{eq:fflow_r1} is rewritten as follows:
\begin{align}
	&\partial_\lambda
	\bp_{\max,\lambda}[\rhoa](\br_1)
	\notag
	\\
	=&
	\frac{1}{2}
	\int_{\br,\br'}
	\partial_\lambda
	\beta v_{\lambda}(\br-\br')
	\notag
	\\
	&\times
	\int_{\br''}
	\frac{\delta^2 \beta F_\lambda[\rhoa]}{\delta \rhoa(\br_1)\delta \rhoa(\br'')}
	\left(
	\rho^{(3)}_\lambda[\rhoa](\br,\br',\br'')
	+
	\rho^{(2)}_\lambda[\rhoa](\br,\br')
	\left[
	\sum_{k=1}^{m}
	\delta(\br-\br'')
	+
	\delta(\br'-\br'')
	-
	\rhoa(\br'')
	\right]	
	\right).
\end{align}
Multiplying the inverse of the second derivative of $\beta F_\lambda[\rhoa]$
and using Eq.~\eqref{eq:finv_r2}, we have
\begin{align}
	\label{eq:rho1_flow}
	0	
	=&
	\int_{\br}
	\left[
	\rho^{(2)}_\lambda[\rhoa](\br_1,\br)
	-
	\rhoa(\br)\rhoa(\br_1)
	\right]
	\partial_\lambda \bp_{\max,\lambda}[\rhoa](\br)
	+
	\rhoa(\br_1)\partial_\lambda \bp_{\max,\lambda}[\rhoa](\br_1)
	\notag
	\\
	&
	-
	\int_{\br}
	\partial_\lambda
	\left[\beta v_\lambda(\br-\br_i)\right]
	\rho^{(2)}_\lambda[\rhoa](\br, \br_1)
	-
	\frac{1}{2}
	\int_{\br,\br'}
	\partial_\lambda\left[\beta v_\lambda(\br-\br')\right]
	\left[
	\rho^{(3)}_\lambda[\rhoa](\br,\br',\br_1)
	-
	\rhoa(\br_1)
	\rho^{(2)}_\lambda[\rhoa](\br,\br')
	\right].
\end{align}
This equation corresponds to Eq.~\eqref{eq:rhom_assume} for $k=1$.

Now we assume that Eq.~\eqref{eq:rhom_assume} holds for all $k\leq m$
and show that this equation also holds for $k=m+1$
by differentiating Eq.~\eqref{eq:rhom_assume}
for $k=m$.
As one can see from Eq.~\eqref{eq:rhom_r},
there appears $\partial_\lambda\delta^2 \beta F_\lambda[\rhoa]/\delta \rhoa(\br)\delta \rhoa(\br')$
from the $\rhoa$-derivative of the first term
in the left-hand side of Eq.~\eqref{eq:rhom_assume}.
Such a factor also appears from the 
$\rhoa$-derivative of $\bp_{\max,\lambda}[\rhoa](\br)$
as one can see from Eq.~\eqref{eq:F_psimax}.
These terms involving the factor 
$\partial_\lambda \delta^2 \beta F_\lambda[\rhoa]/\delta \rhoa(\br)\delta \rhoa(\br')$
cancel each other. Then the $\rhoa$-derivative
of Eq.~\eqref{eq:rhom_assume} for $k=m$ reads
\begin{align}
	&
	\int_{\br'}
	\frac{\delta^2 \beta F_\lambda[\rhoa]}{\delta \rhoa(\br'')\delta \rhoa(\br')}
	\left(
	\partial_\lambda
	\rho^{(m+1)}_\lambda[\rhoa]({\bf r}_1,\cdots,{\bf r}_{m},\br')
	+
	\partial_\lambda
	\rho^{(m)}_\lambda[\rhoa]({\bf r}_1,\cdots,{\bf r}_m)
	\left[
	\sum_{k=1}^{m}\delta({\bf r}_k-{\bf r}')
	-
	\rhoa({\bf r}')
	\right]	
	\right)
	\notag
	\\
	&
	+
	\sum_{i<j}^{m}
	\partial_\lambda
	\left[\beta v_\lambda(\br_i-\br_j)\right]
	\frac{\delta}{\delta\rhoa(\br'')}
	\rho_\lambda^{(m)}[\rhoa](\br_1,\ldots,\br_m)
	\notag
	\\
	=
	&
	\int_{\br}
	\left[
	\frac{\delta}{\delta\rhoa(\br'')}
	\rho^{(m+1)}_\lambda[\rhoa](\br_1,\ldots,\br_m,\br)
	-
	\rhoa(\br)
	\frac{\delta}{\delta\rhoa(\br'')}
	\rho^{(m)}_\lambda[\rhoa](\br_1,\ldots,\br_m)
	\right]
	\partial_\lambda \bp_{\max,\lambda}[\rhoa](\br)
	\notag
	\\
	&
	-
	\rho^{(m)}_\lambda[\rhoa](\br_1,\ldots,\br_m)
	\partial_\lambda \bp_{\max,\lambda}[\rhoa](\br'')
	+
	\frac{\delta}{\delta\rhoa(\br'')}
	\rho^{(m)}_\lambda[\rhoa](\br_1,\ldots,\br_m)
	\sum_{i=1}^m
	\partial_\lambda \bp_{\max,\lambda}[\rhoa](\br_i)
	\notag
	\\
	&
	-
	\int_{\br}
	\sum_{i=1}^m
	\partial_\lambda
	\left[\beta v_\lambda(\br-\br_i)\right]
	\frac{\delta}{\delta\rhoa(\br'')}
	\rho^{(m+1)}_\lambda[\rhoa](\br, \br_1, \ldots, \br_m)
	\notag	
	\\
	&
	-
	\frac{1}{2}
	\int_{\br,\br'}
	\partial_\lambda\left[\beta v_\lambda(\br-\br')\right]
	\left[
	\frac{\delta}{\delta\rhoa(\br'')}
	\rho^{(m+2)}_\lambda[\rhoa](\br,\br',\br_1,\ldots,\br_m)
	-
	\rho^{(2)}_\lambda[\rhoa](\br,\br')
	\frac{\delta}{\delta\rhoa(\br'')}
	\rho^{(m)}_\lambda[\rhoa](\br_1,\ldots,\br_m)
	\right]
	\notag	
	\\
	&
	+
	\frac{1}{2}
	\int_{\br,\br'}
	\partial_\lambda\left[\beta v_\lambda(\br-\br')\right]
	\rho^{(m)}_\lambda[\rhoa](\br_1,\ldots,\br_m)
	\frac{\delta}{\delta\rhoa(\br'')}
	\rho^{(2)}_\lambda[\rhoa](\br,\br').
\end{align}
Multiplying 
$\left(\delta^2 \beta F_{\lambda}/\delta \rhoa\delta \rhoa
	\right)^{-1}[\rhoa]$,
we have
\begin{align}
	\label{eq:rhom1_1}
	&
	\partial_\lambda
	\rho^{(m+1)}_\lambda[\rhoa]({\bf r}_1,\cdots,{\bf r}_{m+1})
	+
	\partial_\lambda
	\rho^{(m)}_\lambda[\rhoa]({\bf r}_1,\cdots,{\bf r}_m)
	\left[
	\sum_{k=1}^{m}\delta({\bf r}_k-{\bf r}')
	-
	\rho(\br_{m+1})
	\right]	
	\notag
	\\
	&
	+
	\sum_{i<j}^{m}
	\partial_\lambda
	\left[\beta v_\lambda(\br_i-\br_j)\right]
	R_\lambda^{(m)}[\rhoa](\br_1,\ldots,\br_m;\br_{m+1})
	\notag
	\\
	=
	&
	\int_{\br}
	\left[
	R_\lambda^{(m+1)}[\rhoa](\br_1,\ldots,\br_m,\br;\br_{m+1})
	-
	\rhoa(\br)
	R_\lambda^{(m)}[\rhoa](\br_1,\ldots,\br_m;\br_{m+1})
	\right]
	\partial_\lambda \bp_{\max,\lambda}[\rhoa](\br)
	\notag
	\\
	&
	-
	\int_{\br}
	\left(\frac{\delta^2 \beta F_{\lambda}}{\delta \rhoa\delta \rhoa}
	\right)^{-1}[\rhoa](\br_{m+1},\br)
	\rho^{(m)}_\lambda[\rhoa](\br_1,\ldots,\br_m)
	\partial_\lambda \bp_{\max,\lambda}[\rhoa](\br)
	\notag
	\\
	&
	+
	R_\lambda^{(m)}[\rhoa](\br_1,\ldots,\br_m;\br_{m+1})
	\sum_{i=1}^m
	\partial_\lambda \bp_{\max,\lambda}[\rhoa](\br_i)
	\notag
	\\
	&
	-
	\int_{\br}
	\sum_{i=1}^m
	\partial_\lambda
	\left[\beta v_\lambda(\br-\br_i)\right]
	R_\lambda^{(m+1)}[\rhoa](\br_1,\ldots,\br_m,\br;\br_{m+1})
	\notag	
	\\
	&
	-
	\frac{1}{2}
	\int_{\br,\br'}
	\partial_\lambda\left[\beta v_\lambda(\br-\br')\right]
	\left[
	R_\lambda^{(m+2)}[\rhoa](\br_1,\ldots,\br_m,\br,\br';\br_{m+1})
	-
	\rho^{(2)}_\lambda[\rhoa](\br,\br')
	R_\lambda^{(m)}[\rhoa](\br_1,\ldots,\br_m;\br_{m+1})
	\right]
	\notag	
	\\
	&
	+
	\frac{1}{2}
	\int_{\br,\br'}
	\partial_\lambda\left[\beta v_\lambda(\br-\br')\right]
	\rho^{(m)}_\lambda[\rhoa](\br_1,\ldots,\br_m)
	R_\lambda^{(2)}[\rhoa](\br,\br';\br_{m+1}).
\end{align}
Here, 
$R_\lambda^{(m)}[\rhoa](\br_1,\ldots,\br_m;\br_{m+1})$
is defined by
\begin{align*}
	R_\lambda^{(m)}[\rhoa](\br_1,\ldots,\br_m;\br)
	=
	\int_{\br'}
	\left(\frac{\delta^2 \beta F_{\lambda}}{\delta \rhoa\delta \rhoa}
	\right)^{-1}[\rhoa](\br,\br')
	\frac{\delta}{\delta\rhoa(\br')}
	\rho_\lambda^{(m)}[\rhoa](\br_1,\ldots,\br_m).
\end{align*}
By use of Eq.~\eqref{eq:rhom_r}, we get
\begin{align}
	&R_\lambda^{(m)}[\rhoa](\br_1,\ldots,\br_m;\br_{m+1})
	\notag
	\\
	&
	\quad
	=
	\rho^{(m+1)}_\lambda[\rhoa]({\bf r}_1,\cdots,\br_{m+1})
	+
	\rho^{(m)}_\lambda[\rhoa]({\bf r}_1,\cdots,{\bf r}_m)
	C(\br_{1},\ldots,\br_{m};\br_{m+1}),
	\\
	&R_\lambda^{(m+1)}[\rhoa](\br_1,\ldots,\br_m,\br;\br_{m+1})
	\notag
	\\
	&
	\quad
	=
	\rho^{(m+2)}_\lambda[\rhoa]({\bf r}_1,\cdots,\br_{m+1},\br)
	+
	\rho^{(m+1)}_\lambda[\rhoa]({\bf r}_1,\cdots,{\bf r}_m,\br)
	\left[
	\delta(\br-\br_{m+1})
	+
	C(\br_{1},\ldots,\br_{m};\br_{m+1})
	\right],
	\\
	&R_\lambda^{(m+2)}[\rhoa](\br_1,\ldots,\br_m,\br,\br';\br_{m+1})
	\notag
	\\
	&
	\quad
	=
	\rho^{(m+3)}_\lambda[\rhoa]({\bf r}_1,\cdots,\br_{m+1},\br,\br')
	\notag
	\\
	&
	\qquad
	+
	\rho^{(m+2)}_\lambda[\rhoa]({\bf r}_1,\cdots,{\bf r}_m,\br,\br')
	\left[
	\delta(\br-\br_{m+1})
	+
	\delta(\br'-\br_{m+1})
	+
	C(\br_{1},\ldots,\br_{m};\br_{m+1})
	\right],
\end{align}
with
\begin{align}
	C(\br_{1},\ldots,\br_{m};\br_{m+1})
	=
	\sum_{k=1}^{m}\delta({\bf r}_k-{\bf r}_{m+1})
	-
	\rhoa({\bf r}_{m+1}).
\end{align}
Using this relation and Eq.~\eqref{eq:rhom_assume},
we find the terms involving the factor
$C(\br_{1},\ldots,\br_{m};\br_{m+1})$
cancel each other.
Then Eq.~\eqref{eq:rhom1_1} is rewritten as follows:
\begin{align}
	\label{eq:rhom1_2}
	&
	\partial_\lambda
	\rho^{(m+1)}_\lambda[\rhoa]({\bf r}_1,\cdots,{\bf r}_{m+1})
	+
	\sum_{i<j}^{m}
	\partial_\lambda
	\left[\beta v_\lambda(\br_i-\br_j)\right]
	\rho_\lambda^{(m+1)}[\rhoa](\br_1,\ldots,\br_{m+1})
	\notag
	\\
	=
	&
	\int_{\br}
	\left[
	\rho_\lambda^{(m+2)}[\rhoa](\br_1,\ldots,\br_{m+1},\br)
	-
	\rhoa(\br)
	\rho_\lambda^{(m+1)}[\rhoa](\br_1,\ldots,\br_{m+1})
	\right]
	\partial_\lambda \bp_{\max,\lambda}[\rhoa](\br)
	\notag
	\\
	&
	-
	\int_{\br}
	\left(\frac{\delta^2 \beta F_{\lambda}}{\delta \rhoa\delta \rhoa}
	\right)^{-1}[\rhoa](\br_{m+1},\br)
	\rho^{(m)}_\lambda[\rhoa](\br_1,\ldots,\br_m)
	\partial_\lambda \bp_{\max,\lambda}[\rhoa](\br)
	\notag
	\\
	&
	+
	\rho_\lambda^{(m+1)}[\rhoa](\br_1,\ldots,\br_{m+1})
	\sum_{i=1}^{m+1}
	\partial_\lambda \bp_{\max,\lambda}[\rhoa](\br_i)
	\notag
	\\
	&
	-
	\int_{\br}
	\sum_{i=1}^{m}
	\partial_\lambda
	\left[\beta v_\lambda(\br-\br_i)\right]
	\rho_\lambda^{(m+2)}[\rhoa](\br_1,\ldots,\br_{m+1},\br)
	-
	\sum_{i=1}^m
	\partial_\lambda
	\left[\beta v_\lambda(\br_{m+1}-\br_i)\right]
	\rho_\lambda^{(m+1)}[\rhoa](\br_1,\ldots,\br_{m+1})
	\notag	
	\\
	&
	-
	\frac{1}{2}
	\int_{\br,\br'}
	\partial_\lambda\left[\beta v_\lambda(\br-\br')\right]
	\left[
	\rho_\lambda^{(m+3)}[\rhoa](\br_1,\ldots,\br_{m+1},\br,\br')
	-
	\rho^{(2)}_\lambda[\rhoa](\br,\br')
	\rho_\lambda^{(m+1)}[\rhoa](\br_1,\ldots,\br_{m+1})
	\right]
	\notag	
	\\
	&
	-
	\int_{\br}
	\partial_\lambda\left[\beta v_\lambda(\br-\br_{m+1})\right]
	\rho_\lambda^{(m+2)}[\rhoa](\br_1,\ldots,\br_{m+1},\br)
	\notag	
	\\
	&
	+
	\frac{1}{2}
	\int_{\br,\br'}
	\partial_\lambda\left[\beta v_\lambda(\br-\br')\right]
	\rho^{(m)}_\lambda[\rhoa](\br_1,\ldots,\br_m)
	\notag
	\\
	&\times
	\left[
	\rho^{(3)}_\lambda[\rhoa](\br,\br',\br_{m+1})
	+
	\rho^{(2)}_\lambda[\rhoa](\br,\br')
	\left(
	\delta(\br-\br_{m+1})
	+
	\delta(\br'-\br_{m+1})
	-
	\rhoa(\br_{m+1})
	\right)
	\right].
\end{align}
By use of Eqs.~\eqref{eq:finv_r2}
and \eqref{eq:rho1_flow}, one finds
that the second and last terms in the right-hand side
of Eq.~\eqref{eq:rhom1_2} cancel each other.
Finally, we can rewrite Eq.~\eqref{eq:rhom1_2}
to obtain Eq.~\eqref{eq:rhom_assume} for $k=m+1$.
Therefore, it is proved that
Eq.~\eqref{eq:rhom_assume} holds for arbitrary $k\geq 1$.
By inserting $\rhoa(\br)=\rhoi(\br)$,
Eq.~\eqref{eq:flow_rhom} is obtained.
\end{widetext}


\begin{thebibliography}{50}%
\makeatletter
\providecommand \@ifxundefined [1]{%
 \@ifx{#1\undefined}
}%
\providecommand \@ifnum [1]{%
 \ifnum #1\expandafter \@firstoftwo
 \else \expandafter \@secondoftwo
 \fi
}%
\providecommand \@ifx [1]{%
 \ifx #1\expandafter \@firstoftwo
 \else \expandafter \@secondoftwo
 \fi
}%
\providecommand \natexlab [1]{#1}%
\providecommand \enquote  [1]{``#1''}%
\providecommand \bibnamefont  [1]{#1}%
\providecommand \bibfnamefont [1]{#1}%
\providecommand \citenamefont [1]{#1}%
\providecommand \href@noop [0]{\@secondoftwo}%
\providecommand \href [0]{\begingroup \@sanitize@url \@href}%
\providecommand \@href[1]{\@@startlink{#1}\@@href}%
\providecommand \@@href[1]{\endgroup#1\@@endlink}%
\providecommand \@sanitize@url [0]{\catcode `\\12\catcode `\$12\catcode
  `\&12\catcode `\#12\catcode `\^12\catcode `\_12\catcode `\%12\relax}%
\providecommand \@@startlink[1]{}%
\providecommand \@@endlink[0]{}%
\providecommand \url  [0]{\begingroup\@sanitize@url \@url }%
\providecommand \@url [1]{\endgroup\@href {#1}{\urlprefix }}%
\providecommand \urlprefix  [0]{URL }%
\providecommand \Eprint [0]{\href }%
\providecommand \doibase [0]{http://dx.doi.org/}%
\providecommand \selectlanguage [0]{\@gobble}%
\providecommand \bibinfo  [0]{\@secondoftwo}%
\providecommand \bibfield  [0]{\@secondoftwo}%
\providecommand \translation [1]{[#1]}%
\providecommand \BibitemOpen [0]{}%
\providecommand \bibitemStop [0]{}%
\providecommand \bibitemNoStop [0]{.\EOS\space}%
\providecommand \EOS [0]{\spacefactor3000\relax}%
\providecommand \BibitemShut  [1]{\csname bibitem#1\endcsname}%
\let\auto@bib@innerbib\@empty
\bibitem [{\citenamefont {Hansen}\ and\ \citenamefont
  {McDonald}(2013)}]{han13}%
  \BibitemOpen
  \bibfield  {author} {\bibinfo {author} {\bibfnamefont {J.}~\bibnamefont
  {Hansen}}\ and\ \bibinfo {author} {\bibfnamefont {I.}~\bibnamefont
  {McDonald}},\ }\href {https://books.google.co.jp/books?id=pbJfOUqZVSgC}
  {\emph {\bibinfo {title} {Theory of Simple Liquids: with Applications to Soft
  Matter}}}\ (\bibinfo  {publisher} {Elsevier Science},\ \bibinfo {year}
  {2013})\BibitemShut {NoStop}%
\bibitem [{\citenamefont {Barker}\ and\ \citenamefont
  {Henderson}(1976)}]{bar76}%
  \BibitemOpen
  \bibfield  {author} {\bibinfo {author} {\bibfnamefont {J.~A.}\ \bibnamefont
  {Barker}}\ and\ \bibinfo {author} {\bibfnamefont {D.}~\bibnamefont
  {Henderson}},\ }\href {\doibase 10.1103/RevModPhys.48.587} {\bibfield
  {journal} {\bibinfo  {journal} {Rev. Mod. Phys.}\ }\textbf {\bibinfo {volume}
  {48}},\ \bibinfo {pages} {587} (\bibinfo {year} {1976})}\BibitemShut
  {NoStop}%
\bibitem [{\citenamefont {Kirkwood}\ \emph {et~al.}(1950)\citenamefont
  {Kirkwood}, \citenamefont {Maun},\ and\ \citenamefont {Alder}}]{kir50}%
  \BibitemOpen
  \bibfield  {author} {\bibinfo {author} {\bibfnamefont {J.~G.}\ \bibnamefont
  {Kirkwood}}, \bibinfo {author} {\bibfnamefont {E.~K.}\ \bibnamefont {Maun}},
  \ and\ \bibinfo {author} {\bibfnamefont {B.~J.}\ \bibnamefont {Alder}},\
  }\href {\doibase 10.1063/1.1747854} {\bibfield  {journal} {\bibinfo
  {journal} {The Journal of Chemical Physics}\ }\textbf {\bibinfo {volume}
  {18}},\ \bibinfo {pages} {1040} (\bibinfo {year} {1950})}\BibitemShut
  {NoStop}%
\bibitem [{\citenamefont {Levesque}(1966)}]{lev66}%
  \BibitemOpen
  \bibfield  {author} {\bibinfo {author} {\bibfnamefont {D.}~\bibnamefont
  {Levesque}},\ }\href {\doibase https://doi.org/10.1016/0031-8914(66)90162-5}
  {\bibfield  {journal} {\bibinfo  {journal} {Physica}\ }\textbf {\bibinfo
  {volume} {32}},\ \bibinfo {pages} {1985} (\bibinfo {year}
  {1966})}\BibitemShut {NoStop}%
\bibitem [{\citenamefont {Parola}\ and\ \citenamefont {Reatto}(1985)}]{par85}%
  \BibitemOpen
  \bibfield  {author} {\bibinfo {author} {\bibfnamefont {A.}~\bibnamefont
  {Parola}}\ and\ \bibinfo {author} {\bibfnamefont {L.}~\bibnamefont
  {Reatto}},\ }\href {\doibase 10.1103/PhysRevA.31.3309} {\bibfield  {journal}
  {\bibinfo  {journal} {Phys. Rev. A}\ }\textbf {\bibinfo {volume} {31}},\
  \bibinfo {pages} {3309} (\bibinfo {year} {1985})}\BibitemShut {NoStop}%
\bibitem [{\citenamefont {Parola}\ \emph {et~al.}(1993)\citenamefont {Parola},
  \citenamefont {Pini},\ and\ \citenamefont {Reatto}}]{par93}%
  \BibitemOpen
  \bibfield  {author} {\bibinfo {author} {\bibfnamefont {A.}~\bibnamefont
  {Parola}}, \bibinfo {author} {\bibfnamefont {D.}~\bibnamefont {Pini}}, \ and\
  \bibinfo {author} {\bibfnamefont {L.}~\bibnamefont {Reatto}},\ }\href
  {\doibase 10.1103/PhysRevE.48.3321} {\bibfield  {journal} {\bibinfo
  {journal} {Phys. Rev. E}\ }\textbf {\bibinfo {volume} {48}},\ \bibinfo
  {pages} {3321} (\bibinfo {year} {1993})}\BibitemShut {NoStop}%
\bibitem [{\citenamefont {Parola}\ and\ \citenamefont {Reatto}(1995)}]{par95}%
  \BibitemOpen
  \bibfield  {author} {\bibinfo {author} {\bibfnamefont {A.}~\bibnamefont
  {Parola}}\ and\ \bibinfo {author} {\bibfnamefont {L.}~\bibnamefont
  {Reatto}},\ }\href {\doibase 10.1080/00018739500101536} {\bibfield  {journal}
  {\bibinfo  {journal} {Advances in Physics}\ }\textbf {\bibinfo {volume}
  {44}},\ \bibinfo {pages} {211} (\bibinfo {year} {1995})}\BibitemShut
  {NoStop}%
\bibitem [{\citenamefont {Parola}\ \emph {et~al.}(2008)\citenamefont {Parola},
  \citenamefont {Pini},\ and\ \citenamefont {Reatto}}]{par08}%
  \BibitemOpen
  \bibfield  {author} {\bibinfo {author} {\bibfnamefont {A.}~\bibnamefont
  {Parola}}, \bibinfo {author} {\bibfnamefont {D.}~\bibnamefont {Pini}}, \ and\
  \bibinfo {author} {\bibfnamefont {L.}~\bibnamefont {Reatto}},\ }\href
  {\doibase 10.1103/PhysRevLett.100.165704} {\bibfield  {journal} {\bibinfo
  {journal} {Phys. Rev. Lett.}\ }\textbf {\bibinfo {volume} {100}},\ \bibinfo
  {pages} {165704} (\bibinfo {year} {2008})}\BibitemShut {NoStop}%
\bibitem [{\citenamefont {Parola}\ \emph {et~al.}(2009)\citenamefont {Parola},
  \citenamefont {Pini},\ and\ \citenamefont {Reatto}}]{par09}%
  \BibitemOpen
  \bibfield  {author} {\bibinfo {author} {\bibfnamefont {A.}~\bibnamefont
  {Parola}}, \bibinfo {author} {\bibfnamefont {D.}~\bibnamefont {Pini}}, \ and\
  \bibinfo {author} {\bibfnamefont {L.}~\bibnamefont {Reatto}},\ }\href
  {\doibase 10.1080/00268970902873547} {\bibfield  {journal} {\bibinfo
  {journal} {Molecular Physics}\ }\textbf {\bibinfo {volume} {107}},\ \bibinfo
  {pages} {503} (\bibinfo {year} {2009})}\BibitemShut {NoStop}%
\bibitem [{\citenamefont {Parola}\ and\ \citenamefont {Reatto}(2012)}]{par12}%
  \BibitemOpen
  \bibfield  {author} {\bibinfo {author} {\bibfnamefont {A.}~\bibnamefont
  {Parola}}\ and\ \bibinfo {author} {\bibfnamefont {L.}~\bibnamefont
  {Reatto}},\ }\href {\doibase 10.1080/00268976.2012.666573} {\bibfield
  {journal} {\bibinfo  {journal} {Molecular Physics}\ }\textbf {\bibinfo
  {volume} {110}},\ \bibinfo {pages} {2859} (\bibinfo {year}
  {2012})}\BibitemShut {NoStop}%
\bibitem [{\citenamefont {Salvino}\ and\ \citenamefont {White}(1992)}]{sal92}%
  \BibitemOpen
  \bibfield  {author} {\bibinfo {author} {\bibfnamefont {L.~W.}\ \bibnamefont
  {Salvino}}\ and\ \bibinfo {author} {\bibfnamefont {J.~A.}\ \bibnamefont
  {White}},\ }\href {\doibase 10.1063/1.462791} {\bibfield  {journal} {\bibinfo
   {journal} {The Journal of Chemical Physics}\ }\textbf {\bibinfo {volume}
  {96}},\ \bibinfo {pages} {4559} (\bibinfo {year} {1992})}\BibitemShut
  {NoStop}%
\bibitem [{\citenamefont {White}\ and\ \citenamefont {Zhang}(1993)}]{whi93}%
  \BibitemOpen
  \bibfield  {author} {\bibinfo {author} {\bibfnamefont {J.~A.}\ \bibnamefont
  {White}}\ and\ \bibinfo {author} {\bibfnamefont {S.}~\bibnamefont {Zhang}},\
  }\href {\doibase 10.1063/1.465263} {\bibfield  {journal} {\bibinfo  {journal}
  {The Journal of Chemical Physics}\ }\textbf {\bibinfo {volume} {99}},\
  \bibinfo {pages} {2012} (\bibinfo {year} {1993})}\BibitemShut {NoStop}%
\bibitem [{\citenamefont {White}\ and\ \citenamefont {Zhang}(1995)}]{whi95}%
  \BibitemOpen
  \bibfield  {author} {\bibinfo {author} {\bibfnamefont {J.~A.}\ \bibnamefont
  {White}}\ and\ \bibinfo {author} {\bibfnamefont {S.}~\bibnamefont {Zhang}},\
  }\href {\doibase 10.1063/1.469716} {\bibfield  {journal} {\bibinfo  {journal}
  {The Journal of Chemical Physics}\ }\textbf {\bibinfo {volume} {103}},\
  \bibinfo {pages} {1922} (\bibinfo {year} {1995})}\BibitemShut {NoStop}%
\bibitem [{\citenamefont {Iso}\ and\ \citenamefont {Kawana}(2019)}]{iso19}%
  \BibitemOpen
  \bibfield  {author} {\bibinfo {author} {\bibfnamefont {S.}~\bibnamefont
  {Iso}}\ and\ \bibinfo {author} {\bibfnamefont {K.}~\bibnamefont {Kawana}},\
  }\href {\doibase 10.1093/ptep/pty148} {\bibfield  {journal} {\bibinfo
  {journal} {Prog. Theor. Exp. Phys.}\ }\textbf {\bibinfo {volume} {2019}},\
  \bibinfo {pages} {013A01} (\bibinfo {year} {2019})},\ \Eprint
  {http://arxiv.org/abs/1808.08133} {arXiv:1808.08133 [cond-mat.stat-mech]}
  \BibitemShut {NoStop}%
\bibitem [{\citenamefont {Caillol}(2006)}]{cai06}%
  \BibitemOpen
  \bibfield  {author} {\bibinfo {author} {\bibfnamefont {J.-M.}\ \bibnamefont
  {Caillol}},\ }\href {\doibase 10.1080/00268970600740774} {\bibfield
  {journal} {\bibinfo  {journal} {Molecular Physics}\ }\textbf {\bibinfo
  {volume} {104}},\ \bibinfo {pages} {1931} (\bibinfo {year}
  {2006})}\BibitemShut {NoStop}%
\bibitem [{\citenamefont {Caillol}(2011)}]{cai11}%
  \BibitemOpen
  \bibfield  {author} {\bibinfo {author} {\bibfnamefont {J.-M.}\ \bibnamefont
  {Caillol}},\ }\href {\doibase 10.1080/00268976.2011.621455} {\bibfield
  {journal} {\bibinfo  {journal} {Molecular Physics}\ }\textbf {\bibinfo
  {volume} {109}},\ \bibinfo {pages} {2813} (\bibinfo {year}
  {2011})}\BibitemShut {NoStop}%
\bibitem [{\citenamefont {Lue}()}]{lue15}%
  \BibitemOpen
  \bibfield  {author} {\bibinfo {author} {\bibfnamefont {L.}~\bibnamefont
  {Lue}},\ }\href {\doibase https://doi.org/10.1002/aic.14868} {\bibfield
  {journal} {\bibinfo  {journal} {AIChE Journal}\ }\textbf {\bibinfo {volume}
  {61}},\ \bibinfo {pages} {2985}}\BibitemShut {NoStop}%
\bibitem [{\citenamefont {Wegner}\ and\ \citenamefont
  {Houghton}(1973)}]{weg73}%
  \BibitemOpen
  \bibfield  {author} {\bibinfo {author} {\bibfnamefont {F.~J.}\ \bibnamefont
  {Wegner}}\ and\ \bibinfo {author} {\bibfnamefont {A.}~\bibnamefont
  {Houghton}},\ }\href {\doibase 10.1103/PhysRevA.8.401} {\bibfield  {journal}
  {\bibinfo  {journal} {Phys. Rev. A}\ }\textbf {\bibinfo {volume} {8}},\
  \bibinfo {pages} {401} (\bibinfo {year} {1973})}\BibitemShut {NoStop}%
\bibitem [{\citenamefont {Wilson}\ and\ \citenamefont {Kogut}(1974)}]{wil74}%
  \BibitemOpen
  \bibfield  {author} {\bibinfo {author} {\bibfnamefont {K.~G.}\ \bibnamefont
  {Wilson}}\ and\ \bibinfo {author} {\bibfnamefont {J.}~\bibnamefont {Kogut}},\
  }\href {\doibase https://doi.org/10.1016/0370-1573(74)90023-4} {\bibfield
  {journal} {\bibinfo  {journal} {Phys. Rep.}\ }\textbf {\bibinfo {volume}
  {12}},\ \bibinfo {pages} {75 } (\bibinfo {year} {1974})}\BibitemShut
  {NoStop}%
\bibitem [{\citenamefont {Polchinski}(1984)}]{pol84}%
  \BibitemOpen
  \bibfield  {author} {\bibinfo {author} {\bibfnamefont {J.}~\bibnamefont
  {Polchinski}},\ }\href {\doibase
  https://doi.org/10.1016/0550-3213(84)90287-6} {\bibfield  {journal} {\bibinfo
   {journal} {Nucl. Phys. B}\ }\textbf {\bibinfo {volume} {231}},\ \bibinfo
  {pages} {269 } (\bibinfo {year} {1984})}\BibitemShut {NoStop}%
\bibitem [{\citenamefont {Wetterich}(1993)}]{wet93}%
  \BibitemOpen
  \bibfield  {author} {\bibinfo {author} {\bibfnamefont {C.}~\bibnamefont
  {Wetterich}},\ }\href {\doibase 10.1016/0370-2693(93)90726-X} {\bibfield
  {journal} {\bibinfo  {journal} {Phys. Lett. B}\ }\textbf {\bibinfo {volume}
  {301}},\ \bibinfo {pages} {90} (\bibinfo {year} {1993})}\BibitemShut
  {NoStop}%
\bibitem [{\citenamefont {Berges}\ \emph {et~al.}(2002)\citenamefont {Berges},
  \citenamefont {Tetradis},\ and\ \citenamefont {Wetterich}}]{ber02}%
  \BibitemOpen
  \bibfield  {author} {\bibinfo {author} {\bibfnamefont {J.}~\bibnamefont
  {Berges}}, \bibinfo {author} {\bibfnamefont {N.}~\bibnamefont {Tetradis}}, \
  and\ \bibinfo {author} {\bibfnamefont {C.}~\bibnamefont {Wetterich}},\ }\href
  {\doibase 10.1016/S0370-1573(01)00098-9} {\bibfield  {journal} {\bibinfo
  {journal} {Phys. Rep.}\ }\textbf {\bibinfo {volume} {363}},\ \bibinfo {pages}
  {223} (\bibinfo {year} {2002})}\BibitemShut {NoStop}%
\bibitem [{\citenamefont {Pawlowski}(2007)}]{paw07}%
  \BibitemOpen
  \bibfield  {author} {\bibinfo {author} {\bibfnamefont {J.~M.}\ \bibnamefont
  {Pawlowski}},\ }\href {\doibase 10.1016/j.aop.2007.01.007} {\bibfield
  {journal} {\bibinfo  {journal} {Ann. Phys.}\ }\textbf {\bibinfo {volume}
  {322}},\ \bibinfo {pages} {2831} (\bibinfo {year} {2007})}\BibitemShut
  {NoStop}%
\bibitem [{\citenamefont {Metzner}\ \emph {et~al.}(2012)\citenamefont
  {Metzner}, \citenamefont {Salmhofer}, \citenamefont {Honerkamp},
  \citenamefont {Meden},\ and\ \citenamefont {Sch\"onhammer}}]{met12}%
  \BibitemOpen
  \bibfield  {author} {\bibinfo {author} {\bibfnamefont {W.}~\bibnamefont
  {Metzner}}, \bibinfo {author} {\bibfnamefont {M.}~\bibnamefont {Salmhofer}},
  \bibinfo {author} {\bibfnamefont {C.}~\bibnamefont {Honerkamp}}, \bibinfo
  {author} {\bibfnamefont {V.}~\bibnamefont {Meden}}, \ and\ \bibinfo {author}
  {\bibfnamefont {K.}~\bibnamefont {Sch\"onhammer}},\ }\href {\doibase
  10.1103/RevModPhys.84.299} {\bibfield  {journal} {\bibinfo  {journal} {Rev.
  Mod. Phys.}\ }\textbf {\bibinfo {volume} {84}},\ \bibinfo {pages} {299}
  (\bibinfo {year} {2012})}\BibitemShut {NoStop}%
\bibitem [{\citenamefont {Dupuis}\ \emph {et~al.}(2021)\citenamefont {Dupuis},
  \citenamefont {Canet}, \citenamefont {Eichhorn}, \citenamefont {Metzner},
  \citenamefont {Pawlowski}, \citenamefont {Tissier},\ and\ \citenamefont
  {Wschebor}}]{dup21}%
  \BibitemOpen
  \bibfield  {author} {\bibinfo {author} {\bibfnamefont {N.}~\bibnamefont
  {Dupuis}}, \bibinfo {author} {\bibfnamefont {L.}~\bibnamefont {Canet}},
  \bibinfo {author} {\bibfnamefont {A.}~\bibnamefont {Eichhorn}}, \bibinfo
  {author} {\bibfnamefont {W.}~\bibnamefont {Metzner}}, \bibinfo {author}
  {\bibfnamefont {J.}~\bibnamefont {Pawlowski}}, \bibinfo {author}
  {\bibfnamefont {M.}~\bibnamefont {Tissier}}, \ and\ \bibinfo {author}
  {\bibfnamefont {N.}~\bibnamefont {Wschebor}},\ }\href {\doibase
  https://doi.org/10.1016/j.physrep.2021.01.001} {\bibfield  {journal}
  {\bibinfo  {journal} {Physics Reports}\ } (\bibinfo {year} {2021}),\
  https://doi.org/10.1016/j.physrep.2021.01.001}\BibitemShut {NoStop}%
\bibitem [{\citenamefont {Polonyi}\ and\ \citenamefont {Sailer}(2002)}]{pol02}%
  \BibitemOpen
  \bibfield  {author} {\bibinfo {author} {\bibfnamefont {J.}~\bibnamefont
  {Polonyi}}\ and\ \bibinfo {author} {\bibfnamefont {K.}~\bibnamefont
  {Sailer}},\ }\href {\doibase 10.1103/PhysRevB.66.155113} {\bibfield
  {journal} {\bibinfo  {journal} {Phys. Rev. B}\ }\textbf {\bibinfo {volume}
  {66}},\ \bibinfo {pages} {155113} (\bibinfo {year} {2002})}\BibitemShut
  {NoStop}%
\bibitem [{\citenamefont {Schwenk}\ and\ \citenamefont
  {Polonyi}(2004)}]{sch04}%
  \BibitemOpen
  \bibfield  {author} {\bibinfo {author} {\bibfnamefont {A.}~\bibnamefont
  {Schwenk}}\ and\ \bibinfo {author} {\bibfnamefont {J.}~\bibnamefont
  {Polonyi}},\ }in\ \href
  {http://theory.gsi.de/hirschegg/2004/Proceedings/Schwenk.ps} {\emph {\bibinfo
  {booktitle} {{32nd International Workshop on Gross Properties of Nuclei and
  Nuclear Excitation: Probing Nuclei and Nucleons with Electrons and Photons
  (Hirschegg 2004) Hirschegg, Austria, January 11-17, 2004}}}}\ (\bibinfo
  {year} {2004})\ pp.\ \bibinfo {pages} {273--282},\ \Eprint
  {http://arxiv.org/abs/nucl-th/0403011} {arXiv:nucl-th/0403011} \BibitemShut
  {NoStop}%
\bibitem [{\citenamefont {Kemler}\ and\ \citenamefont {Braun}(2013)}]{kem13}%
  \BibitemOpen
  \bibfield  {author} {\bibinfo {author} {\bibfnamefont {S.}~\bibnamefont
  {Kemler}}\ and\ \bibinfo {author} {\bibfnamefont {J.}~\bibnamefont {Braun}},\
  }\href {\doibase 10.1088/0954-3899/40/8/085105} {\bibfield  {journal}
  {\bibinfo  {journal} {J. Phys. G}\ }\textbf {\bibinfo {volume} {40}},\
  \bibinfo {pages} {085105} (\bibinfo {year} {2013})}\BibitemShut {NoStop}%
\bibitem [{\citenamefont {Rentrop}\ \emph {et~al.}(2015)\citenamefont
  {Rentrop}, \citenamefont {Jakobs},\ and\ \citenamefont {Meden}}]{ren15}%
  \BibitemOpen
  \bibfield  {author} {\bibinfo {author} {\bibfnamefont {J.~F.}\ \bibnamefont
  {Rentrop}}, \bibinfo {author} {\bibfnamefont {S.~G.}\ \bibnamefont {Jakobs}},
  \ and\ \bibinfo {author} {\bibfnamefont {V.}~\bibnamefont {Meden}},\ }\href
  {\doibase 10.1088/1751-8113/48/14/145002} {\bibfield  {journal} {\bibinfo
  {journal} {Journal of Physics A: Mathematical and Theoretical}\ }\textbf
  {\bibinfo {volume} {48}},\ \bibinfo {pages} {145002} (\bibinfo {year}
  {2015})}\BibitemShut {NoStop}%
\bibitem [{\citenamefont {Kemler}\ \emph {et~al.}(2017)\citenamefont {Kemler},
  \citenamefont {Pospiech},\ and\ \citenamefont {Braun}}]{kem17a}%
  \BibitemOpen
  \bibfield  {author} {\bibinfo {author} {\bibfnamefont {S.}~\bibnamefont
  {Kemler}}, \bibinfo {author} {\bibfnamefont {M.}~\bibnamefont {Pospiech}}, \
  and\ \bibinfo {author} {\bibfnamefont {J.}~\bibnamefont {Braun}},\ }\href
  {\doibase 10.1088/0954-3899/44/1/015101} {\bibfield  {journal} {\bibinfo
  {journal} {J. Phys. G}\ }\textbf {\bibinfo {volume} {44}},\ \bibinfo {pages}
  {015101} (\bibinfo {year} {2017})}\BibitemShut {NoStop}%
\bibitem [{\citenamefont {Liang}\ \emph {et~al.}(2018)\citenamefont {Liang},
  \citenamefont {Niu},\ and\ \citenamefont {Hatsuda}}]{lia18}%
  \BibitemOpen
  \bibfield  {author} {\bibinfo {author} {\bibfnamefont {H.}~\bibnamefont
  {Liang}}, \bibinfo {author} {\bibfnamefont {Y.}~\bibnamefont {Niu}}, \ and\
  \bibinfo {author} {\bibfnamefont {T.}~\bibnamefont {Hatsuda}},\ }\href
  {\doibase 10.1016/j.physletb.2018.02.034} {\bibfield  {journal} {\bibinfo
  {journal} {Phys. Lett. B}\ }\textbf {\bibinfo {volume} {779}},\ \bibinfo
  {pages} {436} (\bibinfo {year} {2018})}\BibitemShut {NoStop}%
\bibitem [{\citenamefont {Yokota}\ \emph
  {et~al.}(2019{\natexlab{a}})\citenamefont {Yokota}, \citenamefont {Yoshida},\
  and\ \citenamefont {Kunihiro}}]{yok18}%
  \BibitemOpen
  \bibfield  {author} {\bibinfo {author} {\bibfnamefont {T.}~\bibnamefont
  {Yokota}}, \bibinfo {author} {\bibfnamefont {K.}~\bibnamefont {Yoshida}}, \
  and\ \bibinfo {author} {\bibfnamefont {T.}~\bibnamefont {Kunihiro}},\ }\href
  {\doibase 10.1103/PhysRevC.99.024302} {\bibfield  {journal} {\bibinfo
  {journal} {Phys. Rev. C}\ }\textbf {\bibinfo {volume} {99}},\ \bibinfo
  {pages} {024302} (\bibinfo {year} {2019}{\natexlab{a}})}\BibitemShut
  {NoStop}%
\bibitem [{\citenamefont {Yokota}\ \emph
  {et~al.}(2019{\natexlab{b}})\citenamefont {Yokota}, \citenamefont {Yoshida},\
  and\ \citenamefont {Kunihiro}}]{yok18b}%
  \BibitemOpen
  \bibfield  {author} {\bibinfo {author} {\bibfnamefont {T.}~\bibnamefont
  {Yokota}}, \bibinfo {author} {\bibfnamefont {K.}~\bibnamefont {Yoshida}}, \
  and\ \bibinfo {author} {\bibfnamefont {T.}~\bibnamefont {Kunihiro}},\ }\href
  {\doibase 10.1093/ptep/pty139} {\bibfield  {journal} {\bibinfo  {journal}
  {Prog. Theor. Exp. Phys.}\ }\textbf {\bibinfo {volume} {2019}},\ \bibinfo
  {pages} {011D01} (\bibinfo {year} {2019}{\natexlab{b}})}\BibitemShut
  {NoStop}%
\bibitem [{\citenamefont {{Yokota}}\ and\ \citenamefont
  {{Naito}}(2019)}]{yok19}%
  \BibitemOpen
  \bibfield  {author} {\bibinfo {author} {\bibfnamefont {T.}~\bibnamefont
  {{Yokota}}}\ and\ \bibinfo {author} {\bibfnamefont {T.}~\bibnamefont
  {{Naito}}},\ }\href {\doibase 10.1103/PhysRevB.99.115106} {\bibfield
  {journal} {\bibinfo  {journal} {\prb}\ }\textbf {\bibinfo {volume} {99}},\
  \bibinfo {pages} {115106} (\bibinfo {year} {2019})}\BibitemShut {NoStop}%
\bibitem [{\citenamefont {Yokota}\ and\ \citenamefont {Naito}(2021)}]{yok21}%
  \BibitemOpen
  \bibfield  {author} {\bibinfo {author} {\bibfnamefont {T.}~\bibnamefont
  {Yokota}}\ and\ \bibinfo {author} {\bibfnamefont {T.}~\bibnamefont {Naito}},\
  }\href {\doibase 10.1103/PhysRevResearch.3.L012015} {\bibfield  {journal}
  {\bibinfo  {journal} {Phys. Rev. Research}\ }\textbf {\bibinfo {volume}
  {3}},\ \bibinfo {pages} {L012015} (\bibinfo {year} {2021})}\BibitemShut
  {NoStop}%
\bibitem [{\citenamefont {Yokota}\ \emph {et~al.}(2020)\citenamefont {Yokota},
  \citenamefont {Kasuya}, \citenamefont {Yoshida},\ and\ \citenamefont
  {Kunihiro}}]{yok20}%
  \BibitemOpen
  \bibfield  {author} {\bibinfo {author} {\bibfnamefont {T.}~\bibnamefont
  {Yokota}}, \bibinfo {author} {\bibfnamefont {H.}~\bibnamefont {Kasuya}},
  \bibinfo {author} {\bibfnamefont {K.}~\bibnamefont {Yoshida}}, \ and\
  \bibinfo {author} {\bibfnamefont {T.}~\bibnamefont {Kunihiro}},\ }\href
  {\doibase 10.1093/ptep/ptaa173} {\bibfield  {journal} {\bibinfo  {journal}
  {Prog. Theor. Exp. Phys.}\ }\textbf {\bibinfo {volume} {2021}},\ \bibinfo
  {pages} {013A03} (\bibinfo {year} {2020})}\BibitemShut {NoStop}%
\bibitem [{\citenamefont {Sumi}\ \emph {et~al.}(2016)\citenamefont {Sumi},
  \citenamefont {Maruyama}, \citenamefont {Mitsutake},\ and\ \citenamefont
  {Koga}}]{sum16}%
  \BibitemOpen
  \bibfield  {author} {\bibinfo {author} {\bibfnamefont {T.}~\bibnamefont
  {Sumi}}, \bibinfo {author} {\bibfnamefont {Y.}~\bibnamefont {Maruyama}},
  \bibinfo {author} {\bibfnamefont {A.}~\bibnamefont {Mitsutake}}, \ and\
  \bibinfo {author} {\bibfnamefont {K.}~\bibnamefont {Koga}},\ }\href {\doibase
  10.1063/1.4953191} {\bibfield  {journal} {\bibinfo  {journal} {The Journal of
  Chemical Physics}\ }\textbf {\bibinfo {volume} {144}},\ \bibinfo {pages}
  {224104} (\bibinfo {year} {2016})},\ \Eprint
  {http://arxiv.org/abs/https://doi.org/10.1063/1.4953191}
  {https://doi.org/10.1063/1.4953191} \BibitemShut {NoStop}%
\bibitem [{\citenamefont {Meeron}\ and\ \citenamefont {Siegert}(1968)}]{mee68}%
  \BibitemOpen
  \bibfield  {author} {\bibinfo {author} {\bibfnamefont {E.}~\bibnamefont
  {Meeron}}\ and\ \bibinfo {author} {\bibfnamefont {A.~J.~F.}\ \bibnamefont
  {Siegert}},\ }\href {\doibase 10.1063/1.1669587} {\bibfield  {journal}
  {\bibinfo  {journal} {The Journal of Chemical Physics}\ }\textbf {\bibinfo
  {volume} {48}},\ \bibinfo {pages} {3139} (\bibinfo {year} {1968})},\ \Eprint
  {http://arxiv.org/abs/https://doi.org/10.1063/1.1669587}
  {https://doi.org/10.1063/1.1669587} \BibitemShut {NoStop}%
\bibitem [{\citenamefont {GROUBA}\ \emph {et~al.}(2004)\citenamefont {GROUBA},
  \citenamefont {ZORIN},\ and\ \citenamefont {SEVASTIANOV}}]{gro04}%
  \BibitemOpen
  \bibfield  {author} {\bibinfo {author} {\bibfnamefont {V.~D.}\ \bibnamefont
  {GROUBA}}, \bibinfo {author} {\bibfnamefont {A.~V.}\ \bibnamefont {ZORIN}}, \
  and\ \bibinfo {author} {\bibfnamefont {L.~A.}\ \bibnamefont {SEVASTIANOV}},\
  }\href {\doibase 10.1142/S0217979204023465} {\bibfield  {journal} {\bibinfo
  {journal} {International Journal of Modern Physics B}\ }\textbf {\bibinfo
  {volume} {18}},\ \bibinfo {pages} {1} (\bibinfo {year} {2004})}\BibitemShut
  {NoStop}%
\bibitem [{\citenamefont {Abe}(1959)}]{abe59}%
  \BibitemOpen
  \bibfield  {author} {\bibinfo {author} {\bibfnamefont {R.}~\bibnamefont
  {Abe}},\ }\href {\doibase 10.1143/PTP.21.421} {\bibfield  {journal} {\bibinfo
   {journal} {Progress of Theoretical Physics}\ }\textbf {\bibinfo {volume}
  {21}},\ \bibinfo {pages} {421} (\bibinfo {year} {1959})}\BibitemShut
  {NoStop}%
\bibitem [{\citenamefont {Ichimaru}(1970)}]{ich70}%
  \BibitemOpen
  \bibfield  {author} {\bibinfo {author} {\bibfnamefont {S.}~\bibnamefont
  {Ichimaru}},\ }\href {\doibase 10.1103/PhysRevA.2.494} {\bibfield  {journal}
  {\bibinfo  {journal} {Phys. Rev. A}\ }\textbf {\bibinfo {volume} {2}},\
  \bibinfo {pages} {494} (\bibinfo {year} {1970})}\BibitemShut {NoStop}%
\bibitem [{\citenamefont {Canet}\ \emph {et~al.}(2003)\citenamefont {Canet},
  \citenamefont {Delamotte}, \citenamefont {Mouhanna},\ and\ \citenamefont
  {Vidal}}]{can03}%
  \BibitemOpen
  \bibfield  {author} {\bibinfo {author} {\bibfnamefont {L.}~\bibnamefont
  {Canet}}, \bibinfo {author} {\bibfnamefont {B.}~\bibnamefont {Delamotte}},
  \bibinfo {author} {\bibfnamefont {D.}~\bibnamefont {Mouhanna}}, \ and\
  \bibinfo {author} {\bibfnamefont {J.}~\bibnamefont {Vidal}},\ }\href
  {\doibase 10.1103/PhysRevD.67.065004} {\bibfield  {journal} {\bibinfo
  {journal} {Phys. Rev. D}\ }\textbf {\bibinfo {volume} {67}},\ \bibinfo
  {pages} {065004} (\bibinfo {year} {2003})}\BibitemShut {NoStop}%
\bibitem [{\citenamefont {Archer}\ \emph {et~al.}(2017)\citenamefont {Archer},
  \citenamefont {Chacko},\ and\ \citenamefont {Evans}}]{arc17}%
  \BibitemOpen
  \bibfield  {author} {\bibinfo {author} {\bibfnamefont {A.~J.}\ \bibnamefont
  {Archer}}, \bibinfo {author} {\bibfnamefont {B.}~\bibnamefont {Chacko}}, \
  and\ \bibinfo {author} {\bibfnamefont {R.}~\bibnamefont {Evans}},\ }\href
  {\doibase 10.1063/1.4993175} {\bibfield  {journal} {\bibinfo  {journal} {The
  Journal of Chemical Physics}\ }\textbf {\bibinfo {volume} {147}},\ \bibinfo
  {pages} {034501} (\bibinfo {year} {2017})}\BibitemShut {NoStop}%
\bibitem [{\citenamefont {Brader}\ and\ \citenamefont {Evans}(2002)}]{bra02}%
  \BibitemOpen
  \bibfield  {author} {\bibinfo {author} {\bibfnamefont {J.}~\bibnamefont
  {Brader}}\ and\ \bibinfo {author} {\bibfnamefont {R.}~\bibnamefont {Evans}},\
  }\href {\doibase https://doi.org/10.1016/S0378-4371(02)00506-X} {\bibfield
  {journal} {\bibinfo  {journal} {Physica A: Statistical Mechanics and its
  Applications}\ }\textbf {\bibinfo {volume} {306}},\ \bibinfo {pages} {287 }
  (\bibinfo {year} {2002})},\ \bibinfo {note} {invited Papers from the 21th
  IUPAP International Conference on St atistical Physics}\BibitemShut {NoStop}%
\bibitem [{\citenamefont {Percus}(1982)}]{per82}%
  \BibitemOpen
  \bibfield  {author} {\bibinfo {author} {\bibfnamefont {J.~K.}\ \bibnamefont
  {Percus}},\ }\href {\doibase 10.1007/BF01011623} {\bibfield  {journal}
  {\bibinfo  {journal} {Journal of Statistical Physics}\ }\textbf {\bibinfo
  {volume} {28}},\ \bibinfo {pages} {67} (\bibinfo {year} {1982})}\BibitemShut
  {NoStop}%
\bibitem [{\citenamefont {Wertheim}(1964)}]{wer64}%
  \BibitemOpen
  \bibfield  {author} {\bibinfo {author} {\bibfnamefont {M.~S.}\ \bibnamefont
  {Wertheim}},\ }\href {\doibase 10.1063/1.1704158} {\bibfield  {journal}
  {\bibinfo  {journal} {Journal of Mathematical Physics}\ }\textbf {\bibinfo
  {volume} {5}},\ \bibinfo {pages} {643} (\bibinfo {year} {1964})}\BibitemShut
  {NoStop}%
\bibitem [{\citenamefont {Verlet}(1964)}]{ver64}%
  \BibitemOpen
  \bibfield  {author} {\bibinfo {author} {\bibfnamefont {L.}~\bibnamefont
  {Verlet}},\ }\href {\doibase https://doi.org/10.1016/0031-8914(64)90204-6}
  {\bibfield  {journal} {\bibinfo  {journal} {Physica}\ }\textbf {\bibinfo
  {volume} {30}},\ \bibinfo {pages} {95} (\bibinfo {year} {1964})}\BibitemShut
  {NoStop}%
\bibitem [{\citenamefont {Blaizot}\ \emph {et~al.}(2006)\citenamefont
  {Blaizot}, \citenamefont {M\'endez-Galain},\ and\ \citenamefont
  {Wschebor}}]{bla06}%
  \BibitemOpen
  \bibfield  {author} {\bibinfo {author} {\bibfnamefont {J.-P.}\ \bibnamefont
  {Blaizot}}, \bibinfo {author} {\bibfnamefont {R.}~\bibnamefont
  {M\'endez-Galain}}, \ and\ \bibinfo {author} {\bibfnamefont {N.}~\bibnamefont
  {Wschebor}},\ }\href {\doibase
  https://doi.org/10.1016/j.physletb.2005.10.086} {\bibfield  {journal}
  {\bibinfo  {journal} {Physics Letters B}\ }\textbf {\bibinfo {volume}
  {632}},\ \bibinfo {pages} {571} (\bibinfo {year} {2006})}\BibitemShut
  {NoStop}%
\bibitem [{\citenamefont {Machado}\ and\ \citenamefont {Dupuis}(2010)}]{mac10}%
  \BibitemOpen
  \bibfield  {author} {\bibinfo {author} {\bibfnamefont {T.}~\bibnamefont
  {Machado}}\ and\ \bibinfo {author} {\bibfnamefont {N.}~\bibnamefont
  {Dupuis}},\ }\href {\doibase 10.1103/PhysRevE.82.041128} {\bibfield
  {journal} {\bibinfo  {journal} {Phys. Rev. E}\ }\textbf {\bibinfo {volume}
  {82}},\ \bibinfo {pages} {041128} (\bibinfo {year} {2010})}\BibitemShut
  {NoStop}%
\bibitem [{\citenamefont {Barker}\ and\ \citenamefont
  {Monaghan}(1962)}]{bar62}%
  \BibitemOpen
  \bibfield  {author} {\bibinfo {author} {\bibfnamefont {J.~A.}\ \bibnamefont
  {Barker}}\ and\ \bibinfo {author} {\bibfnamefont {J.~J.}\ \bibnamefont
  {Monaghan}},\ }\href {\doibase 10.1063/1.1732335} {\bibfield  {journal}
  {\bibinfo  {journal} {The Journal of Chemical Physics}\ }\textbf {\bibinfo
  {volume} {36}},\ \bibinfo {pages} {2564} (\bibinfo {year}
  {1962})}\BibitemShut {NoStop}%
\end{thebibliography}

%

\end{document}